\documentclass[aps,prb,amsmath,amssymb,a4paper,reprint,nofootinbib]{revtex4-2}

\usepackage{verbatim}
\usepackage{graphicx}
\usepackage[utf8]{inputenc}
\usepackage{hyperref}
\usepackage{epstopdf}
\usepackage{listings}
\usepackage{xcolor}
\usepackage{array}
\usepackage{soul}
\usepackage{ragged2e}

\newcommand{\be}{\begin{equation}}
\newcommand{\ee}{\end{equation}}

\newcommand{\md}{\mathrm{d}}
\newcommand{\nn}{\nonumber}
\newcommand{\eF}{\epsilon_\mathrm{_F}}
\newcommand{\Fref}[1]{Fig.~\ref{#1}}
\newcommand{\Eqref}[1]{Eq.~(\ref{#1})}

\makeatletter 
    
\renewcommand\onecolumngrid{
\do@columngrid{one}{\@ne}%
\def\set@footnotewidth{\onecolumngrid}
\def\footnoterule{\kern-6pt\hrule width 1.5in\kern6pt}%
}

\renewcommand\twocolumngrid{
        \def\footnoterule{
        \dimen@\skip\footins\divide\dimen@\thr@@
        \kern-\dimen@\hrule width.5in\kern\dimen@}
        \do@columngrid{mlt}{\tw@}
}%

\makeatother    

\begin{document}

\title{On the theory of supermodulation of the superconducting order parameter created by structural supermodulation of apex distance in optimally doped Bi$_2$Sr$_2$CaCu$_2$O$_{8+x}$
}

\author{Albert M. Varonov}
\email{varonov@issp.bas.bg}
    
\author{Todor M. Mishonov}
\email{mishonov@gmail.com}
\affiliation{Georgi Nadjakov Institute of Solid State Physics, Bulgarian Academy of Sciences, 72 Tzarigradsko Chaussee, BG-1784 Sofia, Bulgaria}
  
\date{\today}
\date{2 Mar 2026}

\begin{abstract}
Recently using Scanning Josephson Tunneling Microscopy (SJTM)
in the group of Séamus Davis
a super-modulation of the superconducting order parameter
induced by super-modulation of the distance $\delta$ between planar Cu and apical O  was observed in O’Mahony~\textit{et.~al.}
[On the electron pairing mechanism of copper-oxide high temperature superconductivity,
PNAS~\textbf{119}(37), e2207449119 (2022),
by S.~M.~O’Mahony, W.~Ren, W.~Chen, Y.~X.~Chong, X.~Liu, H.~Eisaki, S.~Uchida, M.~H.~Hamidian, and J.~C.~Séamus Davis].
The authors conclude that:
``concurrence of prediction from strong correlation theory~\dots~
with these observations indicates that~\dots~super-exchange
is the electron pairing mechanism of Bi$_2$Sr$_2$CaCu$_2$O$_{8+x}$.’’
Additionally, the charge transfer energy 
$\mathcal{E}$, probably between O$2p_z$  and Cu$3d_{x^2-y^2}$ levels was studied by SJTM, too.  
In our current theoretical study we use the LCAO approximation,
Hilbert space spanned on 5 atomic orbitals:
Cu$4s$, Cu$3d_{x^2-y^2}$, O$2p_x$,  O$2p_y$,  O$2p_z$.
For the only super-exchange amplitude $J_{sd}$ we use
the Kondo double electron exchange between 
Cu$4s$ and Cu$3d_{x^2-y^2}$ orbitals and
its anti-ferromagnetic sign is determined 
by adjacent to the copper ion in-plane oxygen orbitals.
Within this approximations we have calculated:
``Measured dependence of~\dots~electron-pair density $n_p$ on the displacement $\delta$ of the apical O atoms from the planar Cu atoms’’
depicted in Fig.~5(C) of O’Mahony~\textit{et.~al.} and obtained an acceptable accuracy.
We conclude that the logarithmic derivative
$\mathcal Q_{n_p}\equiv (\delta/n_p)\,\md\, n_p/\md\,\delta$ is the most convenient
dimensionless parameter to solve the 
``concurrence of predictions from-strong-correlation theories
for hole-doped cuprates’’. 
We also discuss that the correlation between the shape of Fermi 
contour and the critical temperature of optimally hole-doped cuprates can be considered as an analogue of the isotope effect of
phonon superconductors.
As a whole, the analyzed SJTM experiment is one of the best confirmations of 
J.~R\"{o}hler 
[J.~R\"{o}hler, Plane dimpling and Cu$4s$ hybridization in
YBa$_2$Cu$_3$O$_{7-x}$, Physica~B: Cond. Matter
\textbf{284}-\textbf{288}, 104 (2000)]
idea that hybridization of Cu4$s$ with conduction band leads
to increasing of $T_c$.
The lack of an alternative explanations
for SJTM data $n_p$ versus $\delta$ and shape-$T_{c,max}$ correlations for the description of the critical temperature of optimally doped cuprates for several decades 
on the background of a simple view gives
a hint that the long sought pairing mechanism has possibly been found and the Kondo exchange interaction as a property of strongly correlated quantum matter deserves further attention in the physics of layered cuprates.
\end{abstract}

\maketitle

\section{Comment on the electron pairing mechanism of cooper-oxide high temperature superconductor. Motivation}
Our work can be considered as an extended comment on
the article by 
S.M.~O’Mahony, Wangping Ren, Weijiong Chen, Yi Xue Chong, Xiaolong Liu, H.~Eisaki, S.~Uchida, M.H.~Hamidian, 
and J.C. Séamus Davis entitled
\emph{On the electron pairing mechanism of cooper-oxide high temperature superconductor}~\cite{Omahony:22}
To explore the hypothesis where a pairing exchange interaction
is hidden in CuO$_2$ plane the authors combine single-electron and electron-pair {Josephson} scanning tunneling microscopy.
In our opinion their main result is the clear correlation
between the electron pair density $n_\mathrm{p}$
and the displacement $\delta$ of the apical oxygen ions
from the planar copper ions~\cite[Fig.~5C]{Omahony:22}.
We follow their conclusion that the microscopic explanation
of this high correlation dependence indicates the 
pairing mechanism of CuO$_2$ superconductors.
The purpose of the present study is to provide and approximative
BCS analysis describing their 
experimentally observed $n_\mathrm{p}(\delta)$ dependence.
Of course, this is only an approximative self-consistent 
consideration but not an exact mathematical proof
of the uniqueness of the solution.
A lateral aim of our solution is to focus attention of the theorists 
for searching of alternative solutions.

\subsection{Introduction and notations of the LCAO approach}

Soon after the discovery of high-$T_c$ superconductivity 
by Bednorz and Mueller~\cite{Bednorz:86}
a consensus started to emerge 
that a pairing mechanism in the CuO$_2$ plane is related to
double electron exchange. 
Simultaneously it became clear that the electron processes in the
cuprate plane are adequately described by 
Cu$3d_{x^2-y^2}$, Cu$4s$ and O$2p$ $\sigma$-orbitals~\cite{Andersen:95,Andersen:96}.
In each elementary cell $\mathbf{n}=(n_x,\,n_y)$ with
$n_x,\,n_y=0,\,1,\,2,\,\dots, N_x=N_y\gg1$ we have
one Cu ion with $\mathbf{R}_\mathrm{Cu}=(0,\,0,\,0)$,
two in-plane O ions with coordinates
$\mathbf{R}_x=\frac12(1,\, 0,\,0) a_0$
and
$\mathbf{R}_y=\frac12(0,\, 1,\,0)  a_0$,
and one apex oxygen with 
$\mathbf{R}_z=(0,\, 0,\,1)\delta$.
The Linear Combination of Atomic Orbitals (LCAO)
Hamiltonian written in second quantization notations
\begin{widetext}
\begin{align}
\hat H_\mathrm{LCAO} & \! =  \! \sum_{\mathbf{n},\alpha} 
\left\{\hat D_{\mathbf{n},\alpha}^\dagger
\left[-t_{pd}( -\hat X_{\mathbf{n},\alpha}
+\hat X_{x-1,y,\alpha}
+\hat Y_{\mathbf{n},\alpha}
-\hat Y_{x,y-1,\alpha})+\epsilon_d\hat D_{\mathbf{n},\alpha} \right] 
\right.
\label{4_band_LCAO}
\\
& 
+\hat S_{\mathbf{n},\alpha}^\dagger
[-t_{sp} (-\hat X_{\mathbf{n},\alpha}+\hat X_{x-1,y,\alpha} 
-\hat Y_{\mathbf{n},\alpha}+\hat Y_{x,y-1,\alpha})
+t_{az} \hat Z_{\mathbf{n},\alpha}
+\epsilon_s\hat S_{\mathbf{n},\alpha}]
\nn \\
& 
+\hat X_{\mathbf{n},\alpha}^\dagger
[-t_{pp} (\hat Y_{\mathbf{n},\alpha}\! - \! \hat Y_{x+1,y,\alpha} 
\! -\! \hat Y_{x,y-1,\alpha} \! + \! \hat Y_{x+1,y-1,\alpha}) 
\! - \! t_{sp} (-\hat S_{\mathbf{n},\alpha} \! +\! \hat S_{x+1,y,\alpha}) 
\! -\! t_{pd} (-\hat D_{\mathbf{n},\alpha} \! +\! \hat D_{x+1,y,\alpha})
\! +\! \epsilon_p \hat X_{\mathbf{n},\alpha}] \nn \\
&
+\hat Y_{\mathbf{n},\alpha}^\dagger
[-t_{pp} (\hat X_{\mathbf{n},\alpha} \! -\! \hat X_{x-1,y,\alpha} 
\! - \! \hat X_{x,y+1,\alpha} \! +\! \hat X_{x-1,y+1,\alpha}) 
\! - \! t_{sp} (-\hat S_{\mathbf{n},\alpha} \!+ \! \hat S_{x,y+1,\alpha})
\! - \! t_{pd} ( \hat D_{\mathbf{n},\alpha} \! - \! \hat D_{x,y+1,\alpha})
\! +\! \epsilon_p \hat Y_{\mathbf{n},\alpha}]\nn\\
&
+\hat Z_{\mathbf{n},\alpha}^\dagger 
   [t_{as} \hat S_{\mathbf{n},\alpha}
   +\epsilon_{p,z}\hat Z_{\mathbf{n},\alpha}]
\Bigl. \Bigr\},
\nn
\end{align}
\end{widetext}
contains atomic the single-site energies
\be
\epsilon_s>\epsilon_d=0>\epsilon_p\approx\epsilon_{p,z}.
\nn
\ee
and the $\sigma$-transfer integrals between different orbitals
\be
t_{sp}>t_{pd}\gg t_{pp},\qquad 
t_{as}(\delta)= \left(\frac{a_0}{2\delta}\right)^2
\frac{t_{sp}}{Z_a}.
\nn
\ee
As it was emphasized by Anderson~\cite[p. 14, 15, 207]{Anderson_th}
LCAO or Tight Binding (TB) representation is intellectually viable and likely 
very accurate.
For the description of the conduction band we use the physics of
downswings used so successfully by O.~K.~Andersen~\cite[p. 15]{Anderson_th}.
We consider as negligible the O-O hopping integrals 
along the edges of the CuO$_5$ pyramid.
The Fermi operators 
$\hat D_{\mathbf{n},\alpha}$,
$\hat S_{\mathbf{n},\alpha}$,
$\hat X_{\mathbf{n},\alpha}$,
$\hat Y_{\mathbf{n},\alpha}$,
$\hat Z_{\mathbf{n},\alpha}$
denote the annihilation of an electron with spin projection
$\alpha$ in the corresponding orbital in the $\mathbf{n}$-th elementary cell.

Neglecting in first approximation the super-modulation
of the height of the pyramid 
$\delta_\mathbf{n}\approx\mathrm{const}$
the independent electron band structure is given
in complete neglect of differential overlap (CNDO) approximation
by the eigen-value problem of a $5\times5$ matrix
\begin{widetext}
\be
\left(H_\mathrm{LCAO}-\varepsilon_\mathbf{p}\openone\right)\Psi_\mathbf{p}
\equiv
\begin{pmatrix}
\epsilon_d-\varepsilon_\mathbf{p}&0&t_{pd}s_x&-t_{pd}s_y&0\\
0&\epsilon_s^{(0)}-\varepsilon_\mathbf{p}
& t_{sp}s_x&  t_{sp}s_y& t_{as}\\
t_{pd}s_x &t_{sp}s_x&\epsilon_p-\varepsilon_\mathbf{p}
&-t_{pp}s_xs_y&0\\
-t_{pd}s_y&t_{sp}s_y&-t_{pp}s_xs_y&
\epsilon_p-\varepsilon_\mathbf{p}&0\\
0&
t_{as}&0&0&
\epsilon_{p,z}^{(0)}-\varepsilon_\mathbf{p}
\end{pmatrix}
\begin{pmatrix}
D_\mathbf{p}\\
S_\mathbf{p}\\
X_\mathbf{p}\\
Y_\mathbf{p}\\
Z_\mathbf{p}
\end{pmatrix}
=\mathbf{0},
\label{band_Hamiltonian}
\ee
\end{widetext}
where
$\mathbf{p}=(p_x,\,p_y)$ is
2-dimensional (2D) dimensionless momentum
$p_x,\,p_y\in(0,2\pi)$ and
\be
s_x=2\sin(p_x/2),\qquad
s_y=2\sin(p_y/2).
\nn
\ee
The diagonalized independent electron Hamiltonian
\be
\hat H_\mathrm{LCAO}=\sum_{\mathbf{n},b}
(\varepsilon_{\mathbf{p},b}-\epsilon_{_\mathrm{F}})
\hat c_{\mathbf{p},b}^\dagger\hat c_{\mathbf{p},b}
\ee
is also expressed by the Fermi operators
\be
\hat c_{\mathbf{p},b,\alpha}
\hat c_{\mathbf{q},b^\prime,\beta}^\dagger
+\hat c_{\mathbf{q},b^\prime,\beta}^\dagger
\hat c_{\mathbf{p},b,\alpha}
=\delta_{\mathbf{p},\mathbf{q}}
\delta_{\alpha,\beta}\delta_{b,b^\prime}.
\ee
where index $b$ corresponds to band index of every 5 bands
in our 5-orbital consideration.
We have 3 completely filled oxygen bands and
a completely empty Cu$4s$ band.
We have only one conduction Cu$3d_{x^2-y^2}$
band and we omit its band index in our further notations.
In such a way, an electron annihilation field for a Bloch wave
with quasi-momentum $\mathbf{P}=\hbar\mathbf{p}/a_0$
reads 
\begin{align}
\hat \Psi_{\mathbf{p},\alpha}(\mathbf{r})=
\frac{\exp(\mathrm{i}\mathbf{p}\cdot\mathbf{n})}
{\sqrt{N}}
\sum_{\mathbf{n}}
&\,\hat c_{\mathbf{p},\alpha}\left[
S_\mathbf{p}\,
\psi_{\mathrm{Cu}4s}(\mathbf{r}-\mathbf{r}_\mathbf{n})
\right.
\label{LCAO_wave_function}
\\
&
+D_\mathbf{p}\,\psi_{\mathrm{Cu}3d_{x^2-y^2}}
(\mathbf{r}-\mathbf{r}_\mathbf{n})
\nn\\
&+X_\mathbf{p}\,\psi_{\mathrm{O}2p_{x}}(\mathbf{r}-\mathbf{r}_\mathbf{n}-\mathbf{R}_x)\mathrm{e}^{\mathrm{i}\varphi_x}
\nn\\
&+Y_\mathbf{p}\,\psi_{\mathrm{O}2p_{y}}(\mathbf{r}-\mathbf{r}_\mathbf{n}-\mathbf{R}_y)\mathrm{e}^{\mathrm{i}\varphi_y}
\nn\\
&\left. 
+Z_\mathbf{p}\,\psi_{\mathrm{O}2p_{z}}(\mathbf{r}-\mathbf{r}_\mathbf{n}-\mathbf{R}_z)\mathrm{e}^{\mathrm{i}\varphi_z}\right],\nonumber
\end{align}
where
\begin{align}
& N=N_xN_y,
\qquad \varphi_z=0,
\nn\\
&
\varphi_x=\frac12 (p_x-\pi),\quad
\varphi_y=\frac12 (p_y-\pi),\quad
\mathbf{n}=(n_x,\,n_y),
\nn\\
&
n_x=0,1,\dots, N_x\gg1,\qquad
n_y=0,1,\dots, N_y\gg1,\nn\\
&
D_\mathbf{p}^2+S_\mathbf{p}^2
+X_\mathbf{p}^2+Y_\mathbf{p}^2+Z_\mathbf{p}^2=1.
\end{align}
In the present study we use the system of notations pointed out 
by Abrikosov~\cite{Abrikosov:03} as a very clear explanation of the TB method 
applied to CuO$_2$ plane.

Often results of local density approximation (LDA) electron band calculations are criticized that
they give significantly broader conduction band in spite that
the shape of the Fermi contour perfectly matches the ARPES data.
As a compromise we suggest that all energy parameters of the 
LCAO approximation: atomic single site energies $\epsilon_a$,
where $a=s,\,p,\,d$,
inter-atomic transfer amplitudes $t_{ab}$ and Fermi energy
$\eF$ to be divided by a common re-normalizing denominator
\be
\epsilon_a\rightarrow \epsilon_a/Z_\epsilon,\qquad
t_{ab}\rightarrow t_{ab}/Z_\epsilon,\qquad
\eF\rightarrow\eF/Z_\epsilon.
\ee
This change of the energy scale does not change the shape of 
Fermi contour.
The numerical value of the energy scale can be determined from a reliable for the considered material experimental data,
for example the effective mass.
Having a strongly anisotropic, almost 2D
material with a single pocket Fermi contour, all effective masses
are equal:
the cyclotron mass $m_c$, 
the mass related to heat capacity and density of states 
$m_\mathrm{DOS}$,
optical mass $m_\mathrm{opt}$, and
the effective mass of Cooper pairs $m^{*}/2$
\cite{Comment:91}.
For the CuO$_2$ plane the polaronic effects are smaller than the
uncertainties of the experimental methods.
Additionally, for fine tuning we can apply several percent 
correction for the Cu$4s$ apex oxygen O$2p_z$ transfer amplitude $t_{as}\rightarrow t_{as}/Z_a$.


\subsection{BCS gap equation for the \emph{s-d} pairing}

Let us in short juxtapose different double exchange amplitudes in 
the square CuO$_2$ plane.
As it is emphasized in Ref.~\cite{Omahony:22},
the double electron super-exchange 
via oxygen ion between two Cu ions with exchange amplitude 
$J_{dd}\simeq 0.1\,\mathrm{eV}$
is \emph{definitely the mechanism of the CuO$_2$ antiferromagnetic state.}
However, the anti-ferromagnetism of the insulator phase and the superconductivity are not obliged to be created by one and the same double-electron-exchange amplitudes.
Due to uncertainties in the determination of the LCAO parameters
we do not refer to any formula expressing double exchange amplitudes $J$ by the Hubbard $U_{dd}$ 
and the LCAO energy parameters.
For us at least the temporary $J$ is just a parameter of the theory.

The biggest exchange amplitude in the condensed matter physics 
is the Kondo-Zener \emph{s-d} double exchange amplitude $J_{sd}$~\cite{Zener1,Zener2,Zener3,Kondo}.
We recall the Zener double exchange Hamiltonian in Fermi
variables used in the present consideration of 
CuO$_2$ plane~\cite[Eqs.~(2.12) and (4.2)]{MishIndPen:03}
\be
\hat H_\mathrm{Kondo}=-J_{sd}\sum_{\mathbf{n},\,\alpha, \,\beta}
\hat{S}_{\mathbf{n},\,\alpha}^\dagger\hat{D}_{\mathbf{n},\,\beta}^\dagger
\hat{S}_{\mathbf{n},\,\beta}
\hat{D}_{\mathbf{n},\,\alpha},
\label{Zener-Kondo}
\ee
where hats denote Fermi operators
\begin{align}
&
\hat{S}_{\mathbf{n},\,\alpha}
\hat{S}_{\mathbf{m},\,\beta}^\dagger
+\hat{S}_{\mathbf{m},\,\beta}^\dagger
\hat{S}_{\mathbf{n},\,\alpha}
=\delta_{\mathbf{n},\mathbf{m}}
\delta_{\alpha,\beta},\nn\\
&
\hat{D}_{\mathbf{n},\,\alpha}
\hat{S}_{\mathbf{m},\,\beta}^\dagger
+\hat{S}_{\mathbf{m},\,\beta}^\dagger
\hat{D}_{\mathbf{n},\,\alpha}
=0,\nn\\
&
\hat{D}_{\mathbf{n},\,\alpha}
\hat{D}_{\mathbf{m},\,\beta}^\dagger
+\hat{D}_{\mathbf{m},\,\beta}^\dagger
\hat{D}_{\mathbf{n},\,\alpha}
=\delta_{\mathbf{n},\mathbf{m}}
\delta_{\alpha,\beta}.\nn
\end{align}
Zener introduced this double exchange Hamiltonian in order to
explain the ferromagnetism of the iron,
later on Kondo deduced that for some kinetic problems
$J_{sd}$ should to have an anti-ferromagnetic sign~\cite{Kondo}.
This anti-ferromagnetic sign of $J_{sd}$ leads
to a singlet pairing in the CuO$_2$ plane.
The Kondo \emph{s-d} exchange  amplitude
is a function of the Hubbard $U$,
see for example, 
the equation given in
the monograph by 
R.~White~\cite[Eq.~(7.82)]{White:83}.
In this formula even the transition $U\rightarrow\infty$
gives a result for the Kondo amplitude with an anti-ferromagnetic sign.
In such a way, 
one can say that the Kondo interaction is a consequence of the 
strong electron-electron correlations.
The effects of strongly correlated electron systems 
is often reduced to an effective 4-fermion interaction
as performed in \Eqref{Zener-Kondo}.

Such a correlated hopping (or spin exchange) is so
intensive that it has been introduced by Zener several years before the epoch-made BCS paper~\cite{BCS} and now explained in all textbooks on magnetism or review articles on strongly correlated electron systems.
It will be praiseworthy a cluster calculation of $J_{sd}$ for CuO$_2$
to be performed using LDA parameters for single site energies 
and transfer integrals but this task is somehow lateral to our study
in which we consider $J_{sd}$ as phenomenological parameter
of the theory.

The exchange between Cu$4s$ and Cu$3d_{x^2-y^2}$
electrons is performed via 4 in-plane oxygen ligands,
moreover the transfer integral of Cu$4s$ to in-plane
O$2p$ are bigger $t_{sp}>t_{pd}$.
That is why one can expect that $J_{sd}\gg J_{dd}$.
It is quite natural to begin our search of the pairing interaction
with the biggest double exchange amplitude 
introduced several years before the epoch-made BCS article.
In short, we consider the Hamiltonian responsible 
for high-$T_c$ superconductivity (HTS) of the CuO$_2$ plane as
\be
\hat{H}_\mathrm{HTS}
=\hat{H}_\mathrm{LCAO}+\hat{H}_\mathrm{Kondo},
\ee
where $\hat{H}_\mathrm{LCAO}$ is the independent electron Hamiltonian and
$\hat{H}_\mathrm{Kondo}$ is the pairing interaction
created by strong electron-electron correlations
due to big $U_{dd}$.
If any other material $J_{sd}$ has an opposite
ferromagnetic sign, this could lead to a triplet superconductivity
described by a $\hat{H}_\mathrm{s-d}$ exchange interaction.
The generalization for $f$-electrons could be also fruitful.

The calculation of the matrix elements of the Kondo interaction
which participate in the BCS gap equation~\cite[Eq.~(3.6)]{MishIndPen:02}
\be
\Delta_\mathbf{q}=\left<(-V_{\mathbf{q},\mathbf{p}})\,
(\Delta_\mathbf{p}/2E_\mathbf{p})
\tanh(E_\mathbf{p}/2T)
\right>_\mathbf{p}
\label{BCS_gap_equation}
\ee
is a technical algebraic  problem to be described
elsewhere.
In spite created by the environment, the pairing interaction
is located on a single transition ion in the elementary cell.
This leads that the interaction kernel in the BCS integral equation \Eqref{BCS_gap_equation} is separable
\be
V_{\mathbf{p},\mathbf{q}}=-2J_{sd}\,
\chi_{\mathbf{p}}\chi_{\mathbf{q}},
\label{separable}
\ee
as it was postulated in the phenomenological considerations
of the gap anisotropy~\cite{MishIndPen:03,MishIndPen:02}.
For the Kondo interaction analyzed within the LCAO approach,
the separability is simply a property of the microscopic Hamiltonian
$\hat{H}_\mathrm{HTS}$.
We wish also to underline that the
2-electron Kondo exchange amplitude $J_{sd}$
is independent from the apex oxygen and its distance $\delta$.

For the separable pairing interaction \Eqref{separable}
the superconducting gap is factorized by the product of
a temperature dependent order parameter $\Xi(t)$
and the momentum dependent gap anisotropy function
$\chi_\mathbf{p}$ 
\be
\Delta_\mathbf{p}(T)=\Xi(T)\,\chi_\mathbf{p}.
\ee

At these simplifications the BCS gap equation is 
reduced to an algebraic one for the order parameter
$\Xi(T)$~\cite[Eq.~(4.6)]{MishIndPen:02}
\begin{align}
& 2J_{sd}\,
\left<\frac{\chi_\mathbf{p}^2}{2E_\mathbf{p}}
\tanh\left(\frac{E_\mathbf{p}}{2T}\right)\right>_{\!\!\mathrm{p}}
=1,
\label{gap_chi}
\\
&\eta_\mathbf{p}\equiv\varepsilon_\mathbf{p}-\eF,
\qquad \mathbf{p}=a_0 \mathbf{P}/\hbar,
\qquad 
E_\mathbf{p}=\sqrt{\eta_\mathbf{p}^2+\Delta_\mathbf{p}^2},
\nn\\
&\left<F(\mathbf{p})\right>_\mathbf{p}\equiv \int\limits_0^{2\pi}\int\limits_0^{2\pi}
F(p_x,p_y)\,\frac{\md p_x \md p_y}{(2\pi)^2},
\qquad p_x,\,p_y\in (0,\, 2\pi).\nn
\end{align}
When obvious, the index $\mathbf{p}$ will be omitted.
In these notations the density of states at the Fermi level
per spin and CuO$_2$ plaquette reads
\be
\rho_{_\mathrm{F}}=\left<1\,\delta(\eta_\mathbf{p})\right>
\label{rho_F}
\ee
and the averaging of Fermi surface is
\be
\overline{F}=\left<F(\mathbf{p})\delta(\eta_\mathbf{p})\right>
/\rho_{_\mathrm{F}}.
\ee
The analytical results are very convenient for qualitative analysis
but for direct numerical calculations it is simpler to tabulate
the numerical eigen-value and vector problem
\Eqref{band_Hamiltonian} in a grid and
afterwards to interpolate for each argument of the integrand.
Here only the anisotropy gap factor
$\chi_\mathbf{p}^2$ is a new element
in comparison with the classic BCS paper.
Numerically this equation can be solved by successive 
iterations for the order parameter
\be
\Xi^{(n+1)}(T)=\Xi^{(n)}(T)\,
J_{sd}\,
\left<\frac{\chi_\mathbf{p}^2}{E_\mathbf{p}}
\tanh\left(\frac{E_\mathbf{p}}{2T}\right)\right> =1.
\label{Xi(T)}
\ee
On the right side we use the current approximation
$\Xi^{(n)}$ and explicitly calculate the next approximation
$\Xi^{(n+1)}$.
For example, at zero temperature, 
i.e. for $T\ll\Xi(T)\approx\Xi_0$,
we have
\be
\Xi^{(n+1)}_0=J_{sd}
\left<\dfrac{\chi_\mathbf{p}^2}
{\sqrt{\eta_\mathbf{p}^2
+\left(\chi_\mathbf{p}\,\Xi^{(n)}_0\right)^{\!2}}}
\right>
\Xi^{(n)}_0.
\label{T=0_gap_equation}
\ee
The convergence 
\be
\Xi_0=\lim_{n\rightarrow\infty}\Xi_0^{(n)}.
\ee
can be accelerated using the
Wynn epsilon algorithm~\cite{Pade}.
And in such a way we derive $\Xi_0$.
Then we can calculate the critical temperature
\be
T_c\approx\frac{\gamma}{\pi}
\exp\left\{\frac{\left<\chi_\mathbf{p}^2
\ln|\chi_\mathbf{p}|
\delta(\eta_\mathbf{p})\right>}
{\left<\chi_\mathbf{p}^2\delta(\eta_\mathbf{p})\right>}
\right\}
\Xi_0,
\ee
where in the argument of the exponential function
we have some averaged values of the gap anisotropy on the 
Fermi surface 
$\varepsilon_\mathbf{p}=\eF$.
We rewrite this 
Pokrovsky equation in order to get similarity
with the isotropic gap superconductors~\cite{Pokr:61} 
\begin{align} 
&
\frac{2\Xi_0}{T_c}=\frac{2\pi}{\gamma}
\exp\left\{-\frac{\left<\chi_\mathbf{p}^2
\ln|\chi_\mathbf{p}|
\delta(\eta_\mathbf{p})\right>}
{\left<\chi_\mathbf{p}^2\delta(\eta_\mathbf{p})\right>}
\right\},\\
&
\gamma/\pi\approx 0.57,\qquad
\Xi_0=\Xi(T=0)
\qquad
\frac{2\pi}{\gamma}\approx 3.53.
\nn
\end{align}
For further details on the thermodynamics of the anisotropic-gap and multiband clean BCS superconductors see Ref.~\cite{MishonovPenev:02}

The critical temperature $T_c$ can be determined
by \Eqref{Xi(T)} as a point of disappearance of the order parameter
\be
\Xi(T_c)=0.
\ee
Technically knowing that the apex oxygen has a de-pairing influence, 
we can determine the Kondo amplitude at some maximal
pyramid height
\be
\delta_\mathrm{max}\rightarrow T_{c,\mathrm{max}},
\ee
for example $T_{c,\mathrm{max}}=91\,\mathrm{K}$
for $\delta_\mathrm{max}=2.54\,\mathrm{\AA}$
\be
J_{sd}=\dfrac{1}
{\left<\dfrac{\chi_\mathbf{p}^2}{\eta_\mathbf{p}}
\tanh\left(\dfrac{\eta_\mathbf{p}}{2T_{c,\mathrm{max}}}\right)\right>}
\label{J_sd}.
\ee
At known Kondo amplitude $J_{sd}$ we can calculate
from \Eqref{T=0_gap_equation}
the zero-temperature order parameter $\Xi_0$ at
different apex distances $\delta\in[2.30,\,2.58]\,\mathrm{\AA}$.

Qualitatively, the superconducting gap $\Delta$ 
corresponds to a superfluid 
wave function $\Xi\propto\Psi$ and for the density of the superfluid
we have $n_p\propto|\Psi|^2\propto\Xi_0^2$.
For a precise definition we introduce the relative electron-pair density
\be
\tilde{n}_p(\delta)\equiv\frac{\Xi_0^2(\delta)}
{\Xi_0^2(\delta_\mathrm{max})}\le 1.
\label{theory-n_p}
\ee
The dimensionless logarithmic derivative
\be
\mathcal Q_{\tilde{n}_p}=\left.\frac{\delta}{\tilde{n}_p}
\frac{\md \tilde{n}_p}{\md\delta}
\right\vert_{\delta_\mathrm{max}}
\label{Q_theoretical}
\ee
at maximal apex distance can be considered as an
indispensable test for every pairing mechanism 
hidden in the CuO$_2$ plane.
Experimentally in Ref.~\cite{Omahony:22} the electron pair density
is defined to be proportional to the square of the product of
maximal Josephson current $I_J$ and the resistivity of the normal
state $R_N$.
The authors consider that the Josephson voltage $U_J$ is proportional to the superconducting gap $\Delta$, 
to the superconducting order parameter 
$\Xi$ and to the effective wave function $\Psi$~\cite{Omahony:22}.
In such a way, the superfluid density $n_p$ is proportional to the square of modulus of the effective wave function $\Psi$
\begin{align}
&
n_p\propto U_J^2\propto |\Psi|^2,\qquad
U_J\equiv R_NI_J\propto \Delta\propto \Xi \propto \Psi,\\
&
\overline{n}_p(\mathbf{r})\equiv\frac{n_p(\mathbf{r})}{\left< n_p(x,y)\right>_{x,y}},
\end{align}
where brackets in the denominator denote sample surface averaging used for normalization of the electron pair density
$\overline{n}_p$~\cite{Omahony:22}.
One can introduce also the dimensionless logarithmic derivative
\be
\mathcal Q_{\overline{n}_p}\equiv
\left.\frac{\delta}{\overline{n}_p}
\frac{\md \overline{n}_p}{\md\delta}
\right\vert_{\delta_\mathrm{max}}.
\label{Q_experimental}
\ee
We conclude that the logarithmic derivative
$\mathcal Q_{n_p}\equiv (\delta/n_p)\,\md n_p/\md\delta$ is the most convenient
dimensionless parameter to solve the 
concurrence of predictions from-strong-correlation theories
for hole-doped cuprates.

We have performed a
preliminary perturbative analysis using a general
4-band Hamiltonian within the 
\emph{s-d}-LCAO theory supposing that the height of the pyramid
$\delta$ very slowly changes.
Actually, the period of super-modulation 
$\lambda_\mathrm{mod}\sim 26\,\mathrm{\AA}$ is only
twice times larger than the coherence length $\xi_{ab}(0)$ 
for this optimally doped
Bi$_2$Sr$_2$CaCu$_2$O$_{8+x}$ defined as
\be
\left.-T_c\frac{\md B_{c2}(T)}{\md T}
\right\vert_{T_c}
=\frac{\Phi_0}{2\pi\xi_{ab}^2(0)},\quad
\Phi_0=\frac{2\pi\hbar}{|2q_e|},
\ee
where $|2q_e|$ is the Cooper pair charge.
The theory of superconductivity gives the possibility to treat
space in-homogeneous problems and our
$\xi_{ab}(0)/\lambda_\mathrm{mod}<1/2$ 
is only the first step in developing of the complete theory.
Much more important is to find alternative explanations
of the $\overline{n}_p(\delta)$ dependence using preliminary determined parameters.
We must emphasize that
for the calculation of the logarithmic derivative $\mathcal Q_{\overline{n}_p}$ 
of the Cooper pair density
with respect to the apical distance defined in 
\Eqref{Q_experimental},
we not need no fitting parameter. 
We use only LCAO or TB parameters calculated by
\textit{ab initio} LDA within muffin tin potential~\cite{Harrison:80,Andersen:95,Andersen:96}.
We use also the distance dependence of the hopping integral
$t_{sp}\propto1/\delta^2$ and the experimental value of the critical temperature $T_c$.
The energy renormalization which leads to the agreement between
LDA approximation and the ARPES data is not essential for the numerical value
of the calculated value of \Eqref{Q_experimental}.
In short, we use no parameters from the crucial experimenter~\cite{Omahony:22},
we theoretically (without modeling) explain it from first principles.
After this brief review of the \emph{s-d}-LCAO theory,
in the next section we compare our calculation
with the experiment performed in the Davis group
\cite{Omahony:22}.

\section{Solution of the gap equation for different heights $\delta$ of the $\mathrm{CuO_5}$ pyramids. Comparison with the experiment}

\begin{widetext}

\begin{figure}[ht]
\centering
\includegraphics[scale=0.7]{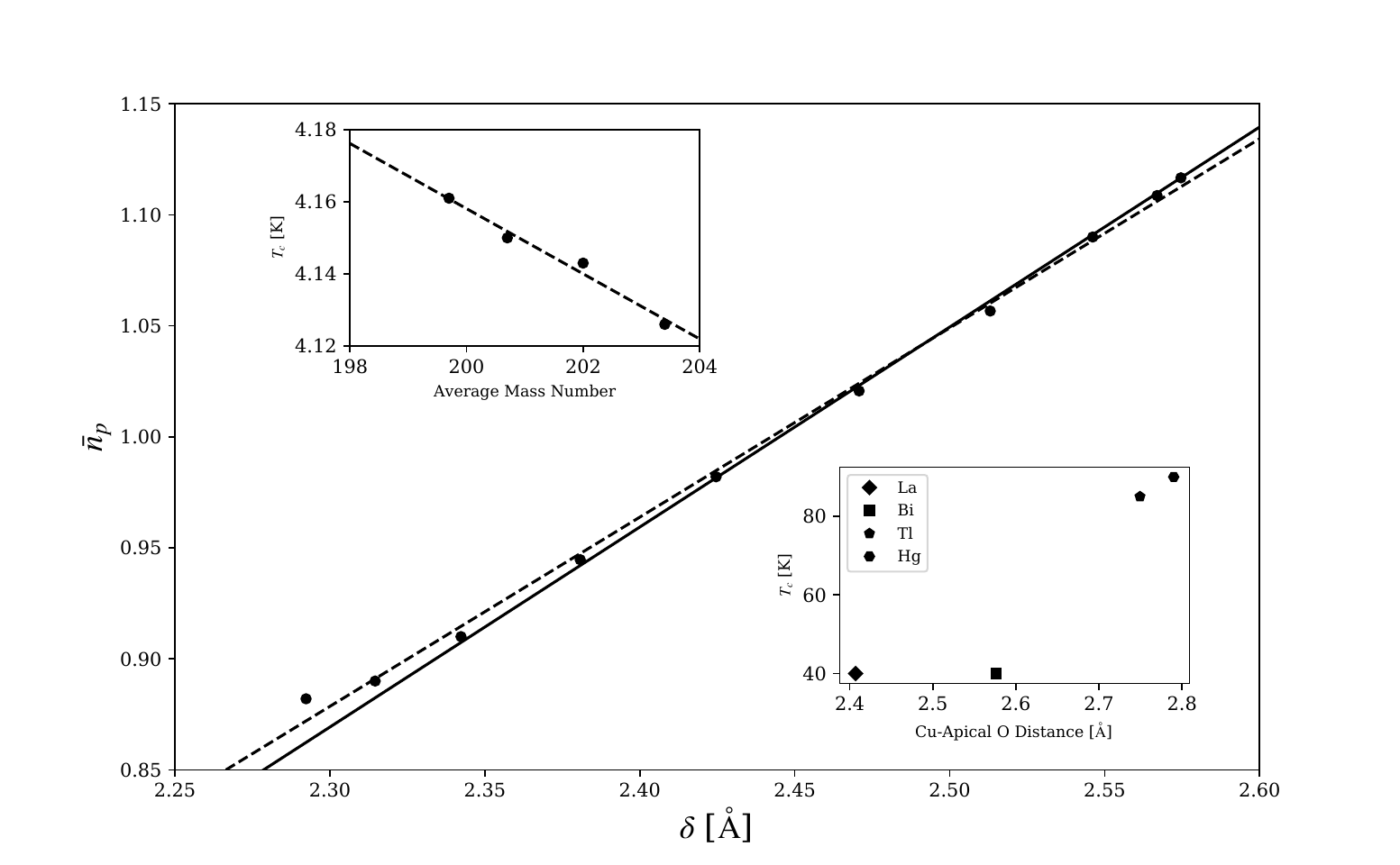}
\caption{
On the \textbf{upper left inset}
the isotope effect for 
mercury~\cite{Reynolds:50} (1950) is reproduced.
The approximately 10\% accuracy of the logarithmic derivative
$Q_{T,M}\equiv \md \ln T_c/\md \ln M\approx\frac12$
triggered the development of phonon mediated pairing theory.
On the \textbf{central large figure}: 
the experimentally determined 
electron pair density $\overline{n}_p$ versus
the apex distance $\delta$ according to
Ref.~\cite[Fig.~5C]{Omahony:22} entitled
\emph{On the electron pairing mechanism of cooper-oxide high
temperature superconductivity.}
The dashed line is the linear regression of the experimental data
with slope approximately defined by \eqref{Q_experimental}.
The continuous line is the theoretical approximation
within the \emph{s-d}-LCAO theory
according to the derivative of the theoretically calculated order parameter according to \eqref{Q_theoretical}.
The \emph{s-d}-LCAO pairing theory gives correct order and evaluation of the slope of the linear regression.
On the \textbf{right lower inset} the correlation $T_c$ versus $\delta$ for different compounds after \cite{Pavarini:01} is given.
The correlation coefficient is much better in $\mathcal{F}$-$\delta$ plot \cite[Fig.~4]{Pavarini:01}
which gives a hint that dependence is via the Cu4s ion
$T_c(\mathcal{F}(\epsilon_s(t_{as}(\delta))))$
as demonstrated by the \emph{s-d}-LCAO pairing theory.
}
\label{Fig:SJTM}
\end{figure}

\end{widetext}

We consider that main result of the SJTM~\cite{Omahony:22}
is the apex distance dependence of the electron pair density
$\overline{n}_p(\delta)\propto (R_NI_J)^2$.
Within the \emph{s-d}-LCAO theory we calculate the 
superfluid density evaluated as
$\tilde{n}_p\propto \Xi^2(\delta)$.
As we analyze the proportionality laws we can fix the scale
of the theoretical calculations at maximal height of the pyramid
determining constant of the proportionality
$C_n$ in order
\be
\tilde{n}_p(\delta_\mathrm{max})
=C_n \overline{n}_p(\delta_\mathrm{max}).
\ee
The results are depicted in \Fref{Fig:SJTM},
one can see two almost coinciding linear dependencies.

For the choice of thes LCAO parameters we followed
Refs.~\onlinecite{Andersen:95,Andersen:96,Pavarini:01}, while the value used for $t_{pp}$ is taken from Ref.~\cite{Mishonov:96}.
The energy re-normalization parameter is chosen
$Z_\epsilon=1.37$ and additionally for the apex hopping $t_{as}$
$Z_a=1.35$ is introduced.
The values of parameters of the theory are given in Table~\ref{tbl:in_energy_mod}
\begin{table}[ht]
\caption{
Single site energies $\epsilon$ and hopping amplitudes $t$ in eV.
The values are taken to be approximate to the ones from
Refs.~\onlinecite{Andersen:95,Andersen:96,Mishonov:96,Pavarini:01}.
}
\begin{tabular}{ c c c c c c c c c c }	
		\hline \hline
		&  \\ [-1em]
		$\epsilon_s$  & $\epsilon_p$  & $\epsilon_d$  &
		$t_{sp}$  & $t_{pp}$\cite{Mishonov:96} & $t_{pd}$  & $f_h$	& $a_0$ & $T_{c,\,\mathrm{max}}$ \\ 
			&  \\ [-1em] 
			4.0	& -0.9 & 0.0 &	2.0	& 0.2 & 1.5 & 0.58 &3.6~\r{A}& 90~K \\
\hline \hline
\end{tabular}
\label{tbl:in_energy_mod}
\end{table}
The hole filling of the conduction band $f_h$ corresponds to optimal doping of $\tilde{p}_\mathrm{opt}=0.16$ holes per CuO$_2$ plaquette 
and its value determines the Fermi energy $\eF$
\begin{align}
f_h(\eF)\equiv
\left<\theta(\varepsilon_\mathbf{p}-\eF)\right>_\mathbf{p}
=\frac{1+\tilde{p}}{2}=1-f_e(\eF),
\end{align}
$f_e$ is the electron filling.
As $\mathbf{p}$ is the dimensionless quasi-momentum,
the wave-vector $\mathbf{k}=\mathbf{p}/a_0$.
The area of the hole pocket in $\mathbf{k}$-space is
\be
S_k=\left(\frac{2\pi}{a_0}\right)^{\!2}f_h,
\qquad
S_P=\left(\frac{2\pi\hbar}{a_0}\right)^{\!2}f_h,
\qquad
\ee
where $S_P$ is the area in the 2D quasi-momentum space
$\mathbf{P}=\hbar\mathbf{p}/a_0.$
According to the Shockley formula~\cite[Eq.~(57.6)]{LL9},
\cite[Eq.~(90.2)]{LL10}, \cite[Eq.~(4.12)]{LifAzbKag},
the energy derivative of this area determines the cyclotron mass
of this almost 2D metal
\be
m_c=\frac{\hbar^2}{2\pi}\left.\frac{\partial S_k}{\partial \epsilon}
\right\vert_{\epsilon=\eF}
=\frac{1}{2\pi}\left.\frac{\partial S_P}{\partial \epsilon}
\right\vert_{\epsilon=\eF}=
2\pi\,\frac{\hbar^2}{a_0^2}\,\rho_{_\mathrm{F}}.
\ee
The same effective mass determines the density of states
$\rho_{_\mathrm{F}}$
\Eqref{rho_F} participating in the heat capacity of the normal phase, for example.

For evaluation of these parameters we calculate
$J_{sd} \approx 5.87$~eV for Fermi energy $\approx 1.37$~eV.
In short: 
(1) the energy scale parameter $Z_e$ is determined by the cyclotron mass $m_c$,
(2) the Kondo exchange amplitude by $T_c(\delta_\mathrm{max})$,
(3) and the apex hopping re-normalization $Z_a$ by
$T_c(\delta_\mathrm{min})$.
This procedure (1, 2, 3) can be repeated until reaching convergence.

\begin{figure}[ht]
\centering
\includegraphics[scale=0.3]{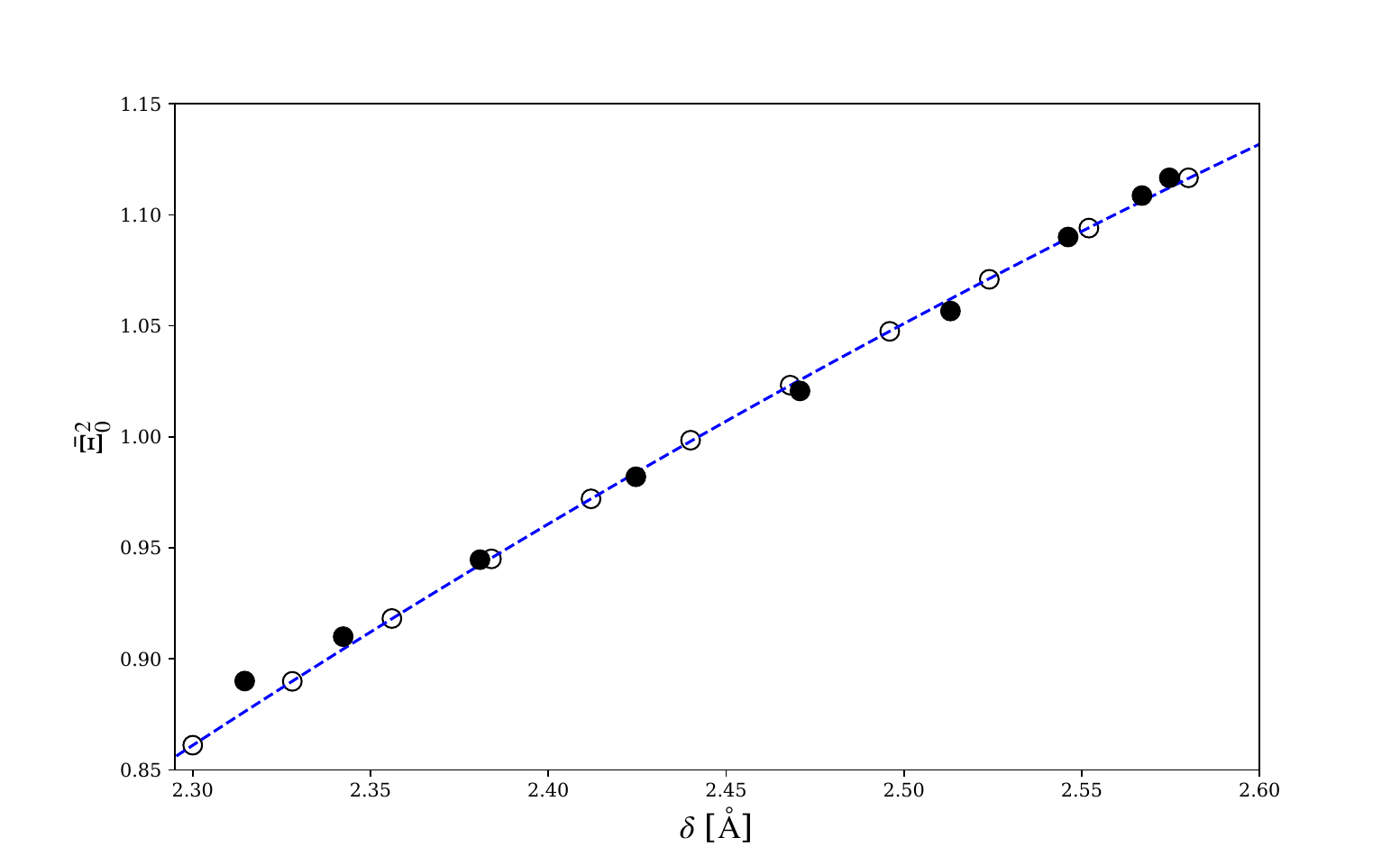}
\caption{
Experimental data from
Ref.~\cite[Fig.~5C]{Omahony:22}
reproduced in \Fref{Fig:SJTM} ($\bullet$)
and our \emph{s-d}-LCAO
calculation according to \Eqref{theory-n_p} ($\circ$).
The theory gives also a small negative curvature (dashed curve based on 2-nd degree polynomial fit) not yet observed by the SJTM.
}
\label{Fig:SJTM-poly}
\end{figure}

An additional analysis reveals that the theory gives a small non-linear deviation shown in \Fref{Fig:SJTM-poly}.
The small negative curvature of the fitting line based on a second degree polynomial fit is easily visible from the numerical calculation based on the theory.

\subsection{\emph{s-d}-LCAO theory in different physical conditions}

The success of one Hamiltonian trying to describe electronic
processes in the CuO$_2$-plane superconductors depends on the diversity of the described phenomena. 

Let us make an analogy with the phonon superconductors.
Aluminum has a very small electron phonon coupling,
it is perfect conductor in the normal state,
however its critical temperature $T_c$ is very small. 
In some sense Al metal is the only one BCS superconductor.
On the other hand, Pb has a significant critical temperature but 
in the normal state due to strong electron-phonon coupling,
it has a very small conductivity and high electron scattering rate.
Is it possible an analogous juxtaposition to be found for high-$T_c$ cuprates?

It is strange but the answer is ``yes’’.
In the normal phase the rate $1/\tau_\mathbf{p}$ of the 
\emph{s-d} exchange electron back-scattering
is determined by the same hybridization function as the gap anisotropy $1/\tau_\mathbf{p}\propto\chi_\mathbf{p}^2$.
This function has zeros along the diagonals of the Brillouin zone
$|p_x|=|p_y|$ and this describes ``cold spot’’ phenomenology
by Ioffe and Milis~\cite{Ioffe:98}
of the scattering rate observed as a narrow electron line in ARPES spectra~\cite{Feng:02}.
One can say that ``cold spots’’ are analogous to the Al metal.
On the other hand, along the lines crossing the hole pocket in
horizontal or vertical direction the gap in the superconducting phase is maximal and in the normal phase the intensive
\emph{s-d} scattering corresponds to the ``hot-spots’’, i.e.
broadening of the ARPES lines,
hence ``hot spots’’ correspond to the Pb metal.
This qualitative consideration together with a conventional 
description of the linear temperature dependence of the resistivity
is discussed in Ref.~\cite[Fig.~3]{hotspot2022}.

For the shape of the Fermi contour, the LCAO approach gives an
excellent agreement to the ARPES experiments or LDA band calculations.
The model equation approximating
the Fermi contour and constant energy curves (CEC) in general
is a simple consequence of the LCAO secular equation
\Eqref{band_Hamiltonian}.
The gap anisotropy 
$\Delta_\mathbf{p}\propto\chi_\mathbf{p}$ 
along the Fermi contour is also in excellent
agreement with the ARPES data
\cite[Fig.~3]{MishIndPen:03}.
The gap anisotropy determines the temperature dependence
of the heat capacity $C(T)$ and the penetration depth
$\lambda_\mathrm{London}(T)$
\cite[Figs.~2, 4]{MishKlenov:05}.
The temperature dependencies $C(T)$, 
$\lambda_\mathrm{London}(T)$ and $\Delta(T)$ was
analyzed in the classical BCS paper.
The agreement with analogous temperature dependencies 
for cuprate superconductors is on the same level and one can say
that in some sense the theory of HTS has already been satisfactory
developed.
The overwhelming correspondence between theory and experiment however, is not yet convincing about the mechanism.

In spite that our interpretation of the experimental 
data~\cite{Omahony:22} differs in some details:
$\mathcal{E}(\delta)$ versus $\epsilon_s(\delta)$,
$J_{dd}(\delta)$ versus $J_{sd}(\delta)$ qualitatively
we arrive at the same conclusion:
the modulation of the apex oxygen distance
modulates the dimensionless BCS coupling constant
$\lambda_{_\mathrm{BCS}}(\delta)$.
For the \emph{s-d} pairing according gap equations
\Eqref{BCS_gap_equation} and
\Eqref{gap_chi} we have
\begin{align}
\lambda_{_\mathrm{BCS}}
(\epsilon_s(t_{as}(\delta_\mathrm{apex})))
& =\left<(-V_{-\mathbf{p},\mathbf{p}}\,
\delta(\varepsilon_\mathbf{p}-\eF)
\right>_\mathbf{p} \\
&=(2J_{sd})\left<\chi_\mathbf{p}^2
\delta(\eta_\mathbf{p})
\right>
=(2J_{sd})\,\overline{\chi_{sd}^2}\,\rho_{_\mathrm{F}}. \nn
\end{align}
The physical meaning of this formula is transparent
-- in order to create Cooper pairs 
it is necessary to average the attraction of charge 
carriers with opposite momenta on the Fermi surface.
The main detail of this formula is the averaged on the 
Fermi surface probability of \emph{s-d} hybridization
\be
\overline{\chi_{sd}^2}(\epsilon_s)
=\frac{\left<\left[S_\mathbf{p}D_\mathbf{p}\right]^2
\delta(\eta_\mathbf{p})
\right>}
{\left<1\, \delta(\varepsilon_\mathbf{p}-\eF)\right>}.
\ee
The dependence of this averaged hybridization probability 
from the position of the Cu$4s$ level $\epsilon_s$
at fixed other LCAO parameters
will be the subject of a separate study.

The modulation of $\epsilon_s$ by the super-modulation
of the apex distance and its corresponding hopping amplitude
$t_{as}(\delta)$ is analogous to a lock-in AC experiment in electronics: the accessories of different chemical compounds 
have been successfully removed. 
In this sense SJTM~\cite{Omahony:22}  
becomes the crucial experiment 
giving the final verdict where the pairing double exchange 
interaction is hidden in the CuO$_2$ plane.
We repeat:
the modulation apex distance $\delta$
simply modulates $\epsilon_s$.
We have given a simple explanation using only formulae 
from the textbooks but will wait for any alternative explanation 
which makes the study of the problem interesting.

Last but not least, we trace the evolution of the idea
on the influence of apex hybridization on the superconductivity properties of the CuO$_2$ plane.
According to the best we know,
the idea that the hybridization of Cu4$s$ with conduction band leads to increasing of $T_c$ was launched by
R\"{o}hler~\cite{Roehler:00,Roehler:00a}
one of these article is entitled 
\emph{Plane dimpling and Cu$4s$ hybridization in
YBa$_2$Cu$_3$O$_{7-x}$}.
Slightly later the idea~\cite{MishIndPen:02}
\emph{
The 3$d$-to-4$s$-by-2$p$ highway to superconductivity in cuprates} was launched
because the Cu$4s$ and Cu$3d_{x^2-y^2}$ are orthogonal
atomic states and their hybridization is possible only
by in-plane O$2p$ states as go-between
if the ligand levels are close transition metal levels.
R\"{o}hler concluded
that the strength of the \emph{spd} hybridization is mainly controlled by the length of apex bound~\cite{Roehler:00,Roehler:00a} and SJTM proved it with a clean experiment~\cite{Omahony:22}.

The apex oxygen raises the Cu$4s$ level,
decreases the hybridization $\chi$
and operates as a de-pairing structural detail.
Qualitatively the de-pairing influence of the apex oxygen was 
observed for different superconductors
in Ref.~\cite[Fig.~4]{Pavarini:01} but without any consideration
of a possible pairing mechanism.

The super-modulation has given the final confirmation
of the influence of the apex oxygen and
we have not many alternative explanations 
for the theoretical guidance on the interplay of 
double electron exchange on the electron pair orders.
For further development one can apply de~Gennes-Bogolyubov
equations to take into account the relatively small value of the period of super-modulation.
Such a final analysis should be accompanied with the derivation
of the formula for the maximal Josephson current
which takes into account the gap anisotropy and 
mutual orientation of sample and tip.

Returning to the commented article~\cite{Omahony:22}
we wish to emphasize
that the authors of this SJTM experimental study 
entitled their article 
by the theoretical achievement related to electron pairing mechanism.
It is unlikely this clean crucial experiment to 
have many alternative explanations: 
like plain vanilla charge transfer phonons with SU(2) symmetry, etc.

• Beyond the experiment that was analyzed, 
let us mark some testable predictions of the presented 
$J_{sd}$-LCAO theory.
\\
\textcolor{blue}{
\begin{enumerate}
\item
In addition to the theory confirmation, there is a prediction: the small negative curvature not observed in the SJTM experiment shown in \Fref{Fig:SJTM-poly}.
\item One and the same set of parameters taken from the explanation of the Josephson voltage super-modulation describe quantitatively the correlation between the shape of the Fermi contour and the critical temperature $T_c$~\cite{PhysB:25}.
\item If we take the slope of the $\overline{n}_p$ versus $\delta$ 
line depicted in \Fref{Fig:SJTM} we have to
estimate $t_{as}$ transfer amplitude between the Cu$4s$ orbital and the apical O$2p_z$ orbital.
This amplitude can be evaluated by tight-binding fit of the electron zone structure.
\item Considering $J_{sd}$ as a main interaction between electrons at least qualitatively, the \textit{s-d} Hamiltonian describes the anisotropy of the ARPES line along the Fermi contour; so called hot-cold spots phenomenology~\cite{hotspot2022}. 
\item Another important part of the theoretical analysis of any crucial experiment is the testable predictions extrapolated theoretically; in general the mission of the theoretical physics is to predict results of an experiment not yet done by anyone.
Concerning the Bi:2212 compound the substitution of Bi by Hg changes the amplitude of the super-modulation $\delta$.
This changes the maximal Josephson current at $T=0$ and the resistivity of the
junction in the normal state $R_N$. 
However, on the basis of the present theory,
we have to point out that we can reliably predict
that the dimensionless logarithmic derivative $\mathcal Q_{\overline{n}_p}$ 
defined in \Eqref{Q_experimental} remains the same within the experimental accuracy;
we need just to wait.
\item Last but not least, a cluster \textit{ab-initio} calculations must confirm that $J_{sd}$ is the most intensive double-electron exchange amplitude exceeding 1~eV.
\end{enumerate}
*) Some initial studies in this direction have already been performed but the final implementation of this schedule should not be the case to published the present final result of the super-modulation of the Josephson voltage and Cooper pair density.
In the present manuscript, a BCS analysis of a contemporary significant experiment is performed, the word `theory’ does not deserve to be written in quotation marks;
it is a matter of professional etiquette.
Putting theory in quotation marks is diminshing end mocking, therefore it is offensive and inappropriate in academic communication. 
}\\

• Despite the topical relation to an experiment, the insight that $T_c$ depends on the apical oxygen distance is not \textit{per se} new to neither the theoretical nor the experimental literature on cuprates.\\
\textcolor{blue}{A: Definitely this idea is from the year of the discovery of high-$T_c$ superconductivity; new is only the theoretical explanation.
In order to emphasize that this is a long standing problem in front of high-$T_c$ superconductivity 
we cite the brilliant idea by J.~R\"{o}hler who concluded
that the strength of the \emph{spd} hybridization is mainly controlled by the length of the apex bond~\cite{Roehler:00,Roehler:00a}.
More than 20 years later, the SJTM experiment~\cite{Omahony:22} confirmed this idea.
In the present manuscript, we represent the indispensable theoretical explanation only.
What means the notion \textit{crucial experiment}, 
this is an experiment giving a clear answer what is the nature of some effect.
Explanation of a clear mechanism which has no any alternatives and giving
the final solution of an old standing problem. 
This is exactly the situation of the SJTM experiment we consider.
The authors of this crucial experiment entitle his article: 
``On the electron pairing mechanism \dots'' because they perfectly understand
that their experiment will have unique explanation leading to final solution 
of a long standing passel.
In our manuscript we represented the theoretical explanation in the framework
of Kondo-Zener exchange interaction incorporated in the standard BCS theory.
The the acceptable small value of parameter 
$\exp(-1/\lambda_{_\mathrm{BCS}})\ll1$ necessary to explain the experimental
data justifies the application of the BCS trial function approach to the problem and 
reveal that influence of strong electron correlations is focused in creation of
4-fermion Kondo exchange interaction parameterized by $J_{sd}$.
A lateral question in this direction is how many crucial experiment related to
CuO$_2$ superconductivity we already have.
How many Hamiltonians (except Kondo one) can explain
the correlation between the shape of Fermi surface and the critical
temperature $T_c$?
The list of crucial experiments can be extended,
hoe many Hamiltonians decribing the gap anisotropy can explain simultaneously
the anisotropy of scattering rate of the normal charge carriers along the Fermy contour above $T_c$. 
}\\
\\
\textcolor{magenta}{
In conclusion we repeat what makes this SJTM experiment~\cite{Omahony:22}
crucial for the revealing of the mechanism of CuO$_2$ high-$T_c$ superconductivity.
The apical O$2p_z$ orbital does not hybridize with Cu$3d_{x^2-y^2}$ and weakly
hybridizes with both planar O$2p_x$ and O$2p_y$ orbitals.
There is a significant hybridization O$2p_z$ has with the Cu$4s$ orbital.
That is why the apical distance modulation modulates the Cu$4s$ energy level.
But this energy level modulation can have so strong influence influence on $T_c$,
only if the Cu$4s$ orbital is involved in the pairing mechanism.
In such a way we finally arrive at the Zener-Kondo \textit{s-d} interaction in which
every Cu$4s$ and Cu$3d_{x^2-y^2}$ exchange their electrons.
In other pairing mechanisms where Cu$4s$ orbital is simply ignored,
it would be impossible \textit{ad hoc} to include the Cu$4s$ orbital in order to reproduce even am order of the magnitude of $\mathcal Q_{\overline{n}_p}$ dimensionless parameter describing the influence of the apical distance on the superfluid density.
In short, we confirm that the statement of 
S.M.~O’Mahony, Wangping Ren, Weijiong Chen, Yi Xue Chong, Xiaolong Liu, H.~Eisaki, S.~Uchida, M.H.~Hamidian, and J.C. Séamus Davis that the JSTM reveals that the electron pairing mechanism of cooper-oxide high temperature superconductors is the long sought crucial confirming experiment~\cite{Omahony:22}.\\
\\
The most important part of the theoretical analysis of any crucial experiment is the testable predictions extrapolated theoretically; in general the mission of the theoretical physics is to predict results of an experiment not yet done by anyone.
Concerning the Bi:2212 compound the substitution of Bi by Hg changes the amplitude of the super-modulation $\delta$.
This changes the maximal Josephson current at $T=0$ and the resistivity of the
junction in the normal state $R_N$. 
However on the basis of the present theory,
we have to point out that we can reliably predict
that the dimensionless logarithmic derivative $\mathcal Q_{\overline{n}_p}$ 
defined in \Eqref{Q_experimental} remains the same within the experimental accuracy;
we need just to wait.
}


\section*{Author contributions}
Both the authors have equally contributed to the writing of the manuscript, programming, making of figures  and experimental data processing.
\\

\section*{Data availability statement}
 The data that support the findings of this study are 
available upon reasonable request from the authors.

\bibliography{Pokrovsky}


\clearpage

\onecolumngrid

\appendix



\onecolumngrid

\clearpage
\thispagestyle{empty}

\appendix

\section*{Editor in Chief of American Physical Society (EIC APS)}
\noindent
10 March 2026\\
Prof. Todor M. Mishonov\\
Institute of Solid State Physics\\
Sofia BG-1784, Bulgaria\\
\\
Re: BJ14878\\
Theory of supermodulation of the superconducting order parameter by supermodulation\\
of the apex distance in optimally doped 
Bi$_2$Sr$_2$CaCu$_2$O$_{8+x}$\\
by Todor M. Mishonov and Albert M. Varonov\\
\\
Dear Prof. Mishonov,\\
\\
I have reviewed the file concerning your manuscript BJ14878 entitled 
``Theory of supermodulation of the superconducting order parameter by supermodulation of the apex distance in optimally doped Bi$_2$Sr$_2$CaCu$_2$O$_{8+x}$''
which was submitted to Physical Review B (PRB). 
The scientific review of your paper is the responsibility
of the editors of the journal. 
The Editor in Chief must assure that the procedures of our journals have been
followed responsibly and fairly in arriving at that decision.\\
\\
Like all of our journals, PRB turns away a fraction of manuscripts without external review. Upon receipt, the editors decided your submission fell into this category. 
You appealed this decision, and the appeal was reviewed by Dr. Titus Neupert, 
a PRB Editorial Board member (EBM), whose recommendation was
against external review.\\
\\
The appeal to the Editor in Chief has to be procedural in nature, that is, 
it does not represent further scientific consideration. 
I judge that proper policy was followed in this review process, 
and further consideration of your submission is not indicated. 
It furthermore seems that this manuscript is a commentary on
[1], https://doi.org/10.1073/pnas.2207449119, 
and such submissions are better directed at the journal that
published the original piece. 
If you have not already contacted PNAS, we recommend you do so.\\
\\
Yours sincerely,\\
\\
Dr. Robert Rosner\\
Editor in Chief\\
American Physical Society\\
\\

\section*{Comment on EIC APS decision}

\begin{itemize}
\item EBM recommendation was against external review.\\
\textcolor{blue}{
As we analyze in the appeal letter to all arguments (``recommendations'')
by EBM was intentionally wrong;
EBM has enough qualification to check the validity of our results.
}
\item The Editor in Chief must assure that the procedures of our journals have been
followed responsibly and fairly in arriving at that decision.\\
\textcolor{blue}{
As it was emphasized in the appeal letter, when all arguments by EBM are intentionally wrong it is already not scientific but a political question of competence of EIC.
EIC officially and responsibly declared that it is fair procedure according the regulations 
of APS.
}
\item Like all of our journals, PRB turns away a fraction of manuscripts without external review. 
Upon receipt, the editors decided your submission fell into this category. 
\textcolor{blue}{Without external and without internal; without any review.
Scientific arguments and scientific achievements are fairly irrelevant factors 
for the journals of APS.
This is the new business model and EBM  \& EIC fconsider that they 
follow the interests of the corporation.
Whether American Physical Society (APS) journals are still academic journals 
is an open question and the public opinion is based in the analysis of many cases analogous to the present one; the detail are given in the next sections.
}

\end{itemize}

\clearpage

\section*{Appeal Letter to APS Editor in Chief}

\noindent
Via Ashot Melikyan, Senior Editor, Physical Review B\\
and PRB Editorial Board Member Prof. Titus Neupert,\\
\\
To APS Editor in Chief,\\
\\
Dear APS Editor in Chief,\\
\\
cc: Colleagues; it is an open letter intended to be publicly accessible:\footnote{\url{https://forbetterscience.com}}\\
Leonid Schneider,
\\

{\centering \textbf{Appeal letter BJ14878 Mishonov}}\\
\\
According to the APS journals editorial policies and practices listed 
\href{https://journals.aps.org/authors/editorial-policies}{here},\footnote{\url{https://journals.aps.org/authors/editorial-policies}}
the scientific evaluation of a manuscript concludes with the advisory opinion of an Editorial Board member.
In our case however, this scientific evaluation is wrong therefore we submit this appeal for broken Physical Review policies and practices.\\
\\
Editorial Board member (EBM) scientific critical points and authors (A):
\begin{itemize}
\item EBM: Beyond the experiment that was analyzed, no testable predictions of the presented ‘theory’ have been pointed out.\\
\textcolor{blue}{A:
This is incorrect, we have provided a tested theoretical explanation of the
numerical amplitude of the observed effect an 
\textcolor{red}{intentionally untrue argument}.}

\item EBM: Ultimately, the connection between the theory and experiment is a near-linear dependence of the electron pair density on the apex distance.
This is obtained while fitting the model parameters such that $T_c$ at the minimal and maximal apex distances found in experiments are matched.\\
\textcolor{blue}{A: This statement may be true for another manuscript but not for ours, as there is no fitting whatsoever there.
This is of course not a misunderstanding because the EBM declares: 
``I have carefully studied the manuscript \dots''.
We arrive at the conclusion that EBM 
\textcolor{red}{intentionally misleads his colleagues editors} with
a wrong statement.
We consider such behavior to be far beyond the professional etiquette.
In our theoretical consideration we use parameters describing electron
band theory but they are taken from the cited in our manuscript 
\textit{ab initio} electron band calculations.
We use also lattice constants, critical temperature, electron charge
but not a single parameter is fitted in order to explain the measured
magnitude of the experimentally observed new effect.}

\item EBM: For this reason it would be imperative to explore 
what the analysis would yield for other possible Hamiltonians (i.e. other pairing mechanisms).\\
\textcolor{blue}{A: It is clear that \textcolor{red}{this imperative by the EBM is inadequate}.
It is a classic imperative to reject a manuscript, why is it necessary to analyse other Hamiltonians when this one provides an excellent result?
The EBM knows this perfectly and that's why he includes this to have grounds for rejection.
If the results were not excellent, that would not be necessary, of course.
Any theoretical work can be rejected explaining a physical effect can be easily rejected by requiring alternative explanations.
We are witnesses of a \textcolor{red}{typical professional misconduct}.
We have a single solution and the EBM will for other Hamiltonians to be tried to explain the effect is like the Mark Twain joke: 
``The knights all married the daughter.  Joy! wassail! finis!.''
\textit{Imperare sibi maximum imperium est.} A Latin sentence. }

\item EBM: The writing style is somewhat colloquial and lengthy -- in line with a comment, but not typical for PRB articles.\\
\textcolor{blue}{A: How is the writing style related to the scientific evaluation of a manuscript?
This cannot be a critical point since nowadays it is a minor issue by easily solvable almost any AI-text generator transforming the writing style to a ``colloquial'' one.
We may even use APS or Physical Review suggested and approved AI generator(s) giving an acceptable writing style for PRB.}

\item EBM: Despite the topical relation to an experiment, 
the insight that $T_c$ depends on the apical oxygen distance is not {\it per se} 
new to neither the theoretical nor the experimental literature on cuprates.\\
\textcolor{blue}{A: Of course this insight is not new but that is not scientifically related to our manuscript; this is not part of our achievements. 
There is a pletora of insights, most of which are unrealized and most probably will stay such forever.
However, this is absolutely irrelevant to our manuscript.
The EBM does not point where this insight has already been realized therefore our study and manuscript are indeed novel.
Therefore, the prase ``not {\it per se} new'' is just an insinuation intended for
subconscious influence on the other editors.
}
\end{itemize}
Having `analyzed' the EBM critical points, it is evident that all of them are erroneous.
And these critical points are quite obviously used for his advisory opinion:\\
\\
``In summary, I see enough weaknesses of the submitted manuscript that I acknowledge the editorial rejection as a justifiable decision.''\\
\\
and\\
\\
``One can of course be of the opinion that any reasonable manuscript written to scientific standards (which the current one certainly is) should undergo review. 
However, this is not anymore the policy of PRB, necessitated by the scarcity of available reviewers.''\\
\\
And we have to reply: The scarcity of available reviewers cannot be the reason the
EBM to write intentionally wrong reviews just to support an initially misunderstanding
of the importance of a manuscript by the Senior Editor of Physical Review B
Dr.~Ashot Melikyan.
And the importance is indisputable, the explained by us experiment was presented at the \href{https://ui.adsabs.harvard.edu/abs/2022APS..MARZ61009O/abstract}{APS March Meeting 2022, abstract id.Z61.009}.\footnote
{\url{https://ui.adsabs.harvard.edu/abs/2022APS..MARZ61009O/abstract},
\url{https://meetings.aps.org/Meeting/MAR22/Session/Z61.9}}
It is not by chance that we are writing this appeal with carbon copy (cc) to the
colleagues which \href{https://ui.adsabs.harvard.edu/abs/2022PNAS..11907449O/citations}{have cited}\footnote{\url{https://ui.adsabs.harvard.edu/abs/2022PNAS..11907449O/citations}} this crucial experiment performed in the group of
Professor~Davis; for them it will be interesting the further development of
the revealing of the mechanism of cuprate high-$T_c$ superconductivity.
This experiment we quantitatively explain has already obtained 
60 citations. 
In such a way, the explanation of the EBM for 
``the scarcity of available reviewers'' is completely inapplicable and inadequate for our manuscript; in our case precisely the opposite statement is true.
Not to mention that any author of the experiment is also a potential peer reviewer.
\\
\\
In addition to this, the scientific ethics has been breached, too.
Putting theory in quotation marks (as \textit{`theory'}) is diminishing, mocking, therefore it is \textcolor{red}{offensive behavior} and inappropriate in academic communication and is against the Physical Review policies and practices, too.
From all this, it is evident that there are no  scientific disagreements with the EBM since he did provide no relevant scientific arguments against our study described in our manuscript that can be found at \href{https://arxiv.org/abs/2506.15726}{arXiv:2506.15726}.\footnote{\url{https://arxiv.org/abs/2506.15726}} \\
\\
In summary, we `see enough weaknesses' of the treatment of our submitted manuscript.
We have given examples of: 
1) offensive behavior,
2) intentionally untrue argument,
3) intentional misleading of other editors 
4) inadequate imperative and
5) professional misconduct.
This certainly is not a fair hearing and we appeal its decision to the APS Editor in Chief.
Young researchers are squeezed to publish articles and have no resources to
defense their honor in case of a bad behavior from the reviewers; 
the obligation of the senior scientists at professor level is at least 
\textbf{for better science}
to maintain an existence minimum of sanitary standards.
\\
\\
Therefore, we ask the APS Editor in Chief to revert our manuscript rejection decision of the PRB editors and to initiate a proper peer review process that will include a proper scientific evaluation.\\
\\
Truly yours,\\
\\
Todor Mishonov and Albert Varonov, 16 December 2025\\
\\
\textit{Post scriptum:}
We have little practice with appeal letters and that is why
we also ask for permission the final incontestable letter to be made public as this open letter.
Thank you in advance for the cooperation.\\
\\
Appendices:\\
1) Letter of PRB editorial board member full professor Dr.~Titus~Neupert and \\
2) Reply to criticism of PRB Editorial Board Member full professor Dr.~T.~Neupert.

\clearpage

\textcolor{blue}{
A: We will repeat some of our arguments in short:\\
\begin{enumerate}
\item
In our study we represent the microscopic theory of a crucial experiment revealing
an important for the physics problem (the mechanism of cuprate high temperature superconductivity).
A EBM (Prof. Dr. Titus Neupert, Full Professor, Condensed Matter Theory)
characterized our manuscript as \textit{`theory'} and quotation marks in this situation
reveal \textcolor{red}{offensive behavior}.
The explained experiment is the quantitative test of our theory.
And we predict that in similar compounds the theory will work with the same
success.
\item
In more detail EBM writes: 
``no testable predictions of the presented ‘theory’ have been pointed out'',
It is not correct we provided a testable theoretical explanation of the
numerical amplitude of the observed effect an 
\textcolor{red}{intentionally untrue argument}.
\item
We emphasize that we are representing first principle behavior but this EBM
writes: ``This is obtained while fitting the model parameters \dots''.
This is of course not a misunderstanding because EBM declares: 
``I have carefully studied the manuscript \dots''.
We arrive in the conclusion that EBM 
\textcolor{red}{intentionally misleads his colleagues editors} with
a wrong statement.
We consider such behavior to be far beyond the professional etiquette.
In our theoretical consideration we use parameters describing electron
band theory but they are taken from the cited in our manuscript 
\textit{ab initio} electron band calculations.
We use also lattice constants, critical temperature, electron charge
but no one parameter is fitted in order to explain the measured
magnitude of the experimentally observed new effect.
\item
What means the notion \textit{crucial experiment}, 
this is an experiment giving a clear answer what is the nature of some effect.
Explanation of a clear mechanism which has no any alternatives and giving
the final solution of an old standing problem. 
This is exactly the situation of the experiment we consider.
The authors of this crucial experiment entitle his article: 
``On the electron pairing mechanism \dots'' because they perfectly understand
that their experiment will have unique explanation leading to final solution 
of a long standing passel.
In our manuscript we represented the theoretical explanation in the framework
of Kondo-Zener exchange interaction incorporated in the standard BCS theory.
What makes EBM in this situation?
He writes: 
\textit{``For this reason it would be imperative to explore 
what the analysis would yield for other possible Hamiltonians (i.e. other pairing mechanisms).''}
We have to reply: the analyzed crucial experiment has unique explanation.
No other Hamiltonians (i.e. other pairing mechanisms) can explain the effect;
for the passed 4 years the experimental work 
has bee well recognized, received many citations, but no one quantitative
theoretical explanation. 
It is clear that \textcolor{red}{the imperative by EBM is inadequate}.
We have a single solution and the EBM will all Hamiltonians to explain the effect
is like Mark Twain joke: 
``The knights all married the daughter.  Joy! wassail! finis!.''
\textit{Imperare sibi maximum imperium est.} A Latin sentence. 
\item We have given the quantitative state-of-the-art theory of a new effect which attracted significant attention. 
We have submitted our manuscript to Physical Review B just because this
crucial experiment has no alternative explanation.
However an EBM requires: \textit{``\dots to explore 
what the analysis would yield for other possible Hamiltonians''}.
Such editorial policy definitely needs of lustration.
Every theoretical work can be rejected because as a rule physical effects 
has unique explanation we are witness of a \textcolor{red}{typical professional misconduct}.
Can EBM suggest alternative explanation of the Schr\"odinger equation describing
hydrogen spectrum. 
With the same success he can find alternative solution of the effect explained in our article.
\item
The EBM imperative [\textit{``For this reason it would be imperative to explore what the analysis would yield for other possible Hamiltonians 
(i.e. other pairing mechanisms).''}] has another aspect:
The experimentalist whose paper we analyze have given to the title of his paper
a theoretical notion: 
[\textit{``On the electron pairing mechanism of copper-oxide high temperature superconductivity''}].
\textcolor{magenta}{
Here we see a serious contradiction:
the authors of the crucial experiment are sure that so subtle effect will
have the unique explanation revealing the mechanism of the whole phenomenon.
But analyzing the quantitative theoretical explanation
(having prejudices to reject the manuscript)
EBM insist the imperative for other possible Hamiltonians
knowing perfectly that it is impossible.
There are precedents for such a behavior but those precedents returns us
century ago in the development of t6he science. 
And the example provided by Prof. Neupert is in such sense obsolete.} 
\item 
The EBM writes:
``One can of course be of the opinion that any reasonable manuscript written to scientific standards (which the current one certainly is) should undergo review. 
However, this is not anymore the policy of PRB,
necessitated by the scarcity of available reviewers.''
And we have to reply: The scarcity of available reviewers can not be the reason
EBM to write intentionally wrong reviews just to support an initially misunderstanding
of the importance of an manuscript by Senior Editor Physical Review B
Dr.~Ashot Melikyan.
\end{enumerate}
}
\textcolor{blue}{A: We have give examples of: 
1) offensive behavior,
2) intentionally untrue argument,
3) intentional misleading of other editors 
4) inadequate imperative as
5) professional misconduct.
We feel that editors are not adhered to the 
Physical Review policies and practices.
We emphasized some items which can have 1 bit evaluation Yes or No.
We analyzed that all arguments pointed out by EBM are wrong and
our paper do nor received a fair hearing.
Young researcher are squeezed to publish articles and have no resources to
defense his honor in case of bad behavior from the reviewers; 
the obligation of senior scientists in professor level is to maintain
at least existence minimum sanitary standards.
}\\
\\
\textcolor{blue}{A: Of course we can be in error, 
but in spite of the scarcity of available reviewers,
due to the the extreme importance of the problem of pairing mechanism of cyprate hight temperature superconductivity,
we took the liberty kindly to ask you for a cooperation:
our manuscript to receive only one bit review:}\\
\textcolor{magenta}{
\textbf{
Whether our study gives an \textit{ab initio} theoretical explanation
of the crucial experiment revealing the mechanism of cuprate
high temperature superconductivity Yes/No.}
}\\
\textcolor{blue}{We have small practice with appeal letter and that is why
we also ask for permission to publish in Internet your final incontestable letter.
Thank you in advance for the cooperation.\\
\\
Truly yours,\\
\\
Todor Mishonov and Albert Varonov, 13 December 2025\\
}
\\
\\
\textbf{Post-Postscript}. 
In order colleagues in Cc to be able to follow the discussion in the present
open letter we are providing irrelevant for the Physical Review
appendices which are accessible in the journal archive related 
to the present manuscript.
In addition if our appeal to the chief editor will lead to a real reviewing to our manuscript eventually trough other EBM we are giving appendices related to the science.\\
\\
Appendices:\\
1) Scientific details to the appeal Letter to APS Editor in Chief \ref{Scientific details}\\
2) Letter of PRB editorial board member by Prof. Neupert
\ref{Neupert_letter} and \\
3) Reply to Editorial Criticism by 
PRB Editorial Board Member Prof. T. Neupert
\ref{Reply_to_Neupert}

\clearpage 

\subsection{Appendix: Scientific details to the appeal Letter to APS Editor in Chief}
\label{Scientific details}
\noindent
Dear APS Editor in Chief, dear Editors, dear colleagues from APS,\\
\\
Sofia, 16 November 2025\\

{\centering \textbf{Appeal letter BJ14878 Mishonov}}\\
\\
Due to the unprecedented importance of the long standing problem of the mechanism of high-temperature superconductivity,
we take the liberty to address the following appeal for consideration of our manuscript to this high level.
Here we cite 2 points from the PRB editorial board member Prof. Titus Neupert:
\\ \\
• in my role as PRB editorial board member, I have been asked to handle the appeal on BJ14878
concerning its editorial rejection. I have carefully studied the manuscript, the relevant Ref.~1, and all
previous correspondence.\\
\\
• Ultimately, the connection between the theory and experiment is a near-linear dependence of the
electron pair density on the apex distance. This is obtained while fitting the model parameters such
that Tc at the minimal and maximal apex distances found in experiments are matched. In any theory,
to lowest order, Tc would depend linearly on any model parameter, unless a linear term is forbidden,
e.g. by symmetry. For this reason it would be imperative to explore what the analysis would yield for
other possible Hamiltonians (i.e. other pairing mechanisms).\\
\\
Here we have to emphasize that we have made an \textit{ab initio} calculation without any fitting parameters.
It was clearly expressed in the submitted manuscript and 
if Prof. Neupert had read and understood the manuscript, he would not have suggested that we fit an almost linear experimental dependence 
with a line passing through the end points.
That is why those two points from the letter by a PRB editorial board member are in contradiction.
The editorial board member has not  ``carefully studied the manuscript’’ as it is mentioned in the first point.
It is ridiculous an important study to be rejected by such type of misunderstanding and carelessness, therefore we kindly ask PRB and APS
to alleviate the choice of an appropriate senior scientists able to read and comprehend a condensed matter physics manuscript.\\
\\
Recently, using Scanning Josephson Tunneling Microscopy (SJTM),
in the group of Séamus Davis
a super-modulation of the superconducting order parameter
induced by super-modulation of the distance $\delta$ between planar Cu and apical O  was observed in
in Bi$_2$Sr$_2$CaCu$_2$O$_{8-x}$ using Scanned Josephson Tunneling Microscopy (SJTM)
O’Mahony~\textit{et.~al.}~[On the electron pairing mechanism of copper-oxide high
temperature superconductivity,
PNAS~\textbf{119}(37), e2207449119 (2022),
by S.~M.~O’Mahony, W.~Ren, W.~Chen, Y.~X.~Chong, X.~Liu, H.~Eisaki, S.~Uchida, M.~H.~Hamidian, and J.~C.~Séamus Davis].
In spite being an experimental study, the authors choose a theoretical title ``electron pairing mechanism\dots’’.
They strongly believe that this experiment is the crucial one revealing the interaction creating superconductivity in
``copper-oxide high temperature superconductivity’’. 
This important article has attracted significant interest and has already received more than 50 citations.
But none of them, according to the best we know is the theoretical explanation.
If that's the case, a proper review must point this.
\\
\\
In our manuscript which we resubmit to be considered for publication in PRB,
we give a state-of-the-art theory of the observed by SJTM effect.
We wish to stress out that our theoretical calculation has no fitting parameters related to explained experiment.
The Bi$_2$Sr$_2$CaCu$_2$O$_{8-x}$ compound is very well studied material with known structure,
critical temperature $T_c$ and electronic band structure with reliable tight binding approximation of the parameters of the conduction plane..\\
\\
In short, we have calculated and adequate dimensionless parameter allowing the theory to be compared with the experiment:
the logarithmic derivative of the cooper pair density as a function of the apical oxygen distance.
We have reached an excellent several percent agreement with the parameter extracted from the experimental data.\\
\\
In short, our theoretical result has lead to the conclusion that this SJTM experiment Ref.~\cite{Omahony:22} is the crucial one for determination of the pairing mechanism in high-$T_c$ superconductivity; in full analogy with the isotope effect in phonon
superconductors many decades ago.\\
\\
Our theory is based on the Tight Binding (TB) approximation of the LDA calculations of the electron band theory,
Kondo-Zener \textit{s-d} exchange interaction which has the biggest double exchange amplitude $J_{sd}$ in condensed matter physics, and the standard BCS theory in which the first two ingredients are incorporated.
In short, we use only well-known formulae described in textbooks published more than 20 years ago.
That is why the validity of our calculation can be checked by any colleague graduated in condensed matter physics.\\
\\
The importance of the solved problem, the possible interest for the PRB readers however, has to be evaluated by 
a senior physicist actively working in condensed matter theoretical physics.
The importance and the interest for the physics are created by the citations of the SJTM article~\cite{Omahony:22}\\
\\
The reviewer has to start only with one bit answer related to the accusation of fitting.
The manuscript has no fitting parameters: Yes/No.
And if yes, which are these, of course?\\
\\
We strongly believe that the prejudices related to the journal business model can be skipped in favor of an academic consideration of the solution of one important problem;
PRB is still and academic and scientific journal (hopefully).\\
\\
We analyzed some points of disagreement with the editorial board member.
They are all in error and in some moments intentionally insulting.
But in order to solve the general disagreement, let us focus on the most typical example.
The editorial board member writes:
``This is obtained while fitting the model parameters such that $T_c$ at the minimal and maximal apex distances found in experiments are matched.
In any theory, to lowest order, $T_c$ would depend linearly on any model parameter, unless a linear term is forbidden, e.g. by symmetry.''\\
\\
We appreciate the lesson on analytical geometry that between two points only
one straight line can be drawn. 
However we must emphasize that {\em there are no fitting parameters in our study.
and no single parameter is taken from the experimental study which we explain}.
Our theoretical study is based on the parameters of 
high-$T_c$ superconductivity published long time ago in leading physical journals.
The BCS approximation $\exp(-1/\lambda_\mathrm{BCS})$ is an acceptable treatment when $T_c\ll$ the band width.
If for all questions the reply is ``\textbf{Yes}'',  we consider that our manuscript deserves `further' professional reviewing.\\
\\
It is ridiculous a manuscript to be discriminated by lack or referee reports
only because all arguments by the editors are erroneous and irrelevant.
As appendix to this letter we include our point by point analysis of the letter
(actually report) PRB editorial board member by Prof. Neupert.
\\
\\
\textbf{In conclusion, on the basis of the good scientific practice, we are kindly asking our theoretical analysis of the crucial experiment of revealing the mechanism high-$T_c$ in cuprates to be considered for publication in PRB by professionally working in the field colleagues able to trace application of Kondo-Zener exchange Hamiltonian
into BCS theory; we are not asking about the Moon.
If necessary we can resubmit an amended version of the manuscript in which in color
we have inserted the main details from the present appeal without any
change of the scientific content simply because it's neither necessary nor required.
}
\\
\\
Truly yours,\\
\\
Todor Mishonov and Albert Varonov
\\
\\
Appendices:\\
2) Review by PRB Editorial Board Member Prof. Titus Neupert
\ref{Neupert_letter} and \\
3) Reply to Editorial Criticism by PRB Editorial Board Member Prof. T. Neupert
\ref{Reply_to_Neupert}

\clearpage

\subsection{Review by PRB Editorial Board Member Prof. Titus Neupert}
\label{Neupert_letter}
\noindent
\\
Appeal letter BJ14878 Mishonov\\
\\
Dear Dr. Melikyan, dear authors of BJ14878,\\
\\
in my role as PRB editorial board member, I have been asked to handle the appeal on BJ14878
concerning its editorial rejection. I have carefully studied the manuscript, the relevant Ref.~1, and all
previous correspondence.\\
\\
For an appeal to be successful, it must be clear that the standard editorial policies and practices have
not been duly followed or the manuscript has been clearly misjudged (in the case at hand: by the
editors). The former is not the case here, as editorial rejections are standard practice in PRB and at
the discretion of the editors.\\
\\
What speaks in favor of the manuscript is that is presents a quantitative theoretical analysis on a relevant experiment and an outstanding 
open problem in theoretical physics.\\
\\
Critical points are:\\
\\
• Beyond the experiment that was analyzed, no testable predictions of the presented ‘theory’ have
been pointed out.\\
\\
• Ultimately, the connection between the theory and experiment is a near-linear dependence of the
electron pair density on the apex distance. This is obtained while fitting the model parameters such
that Tc at the minimal and maximal apex distances found in experiments are matched. In any theory,
to lowest order, Tc would depend linearly on any model parameter, unless a linear term is forbidden,
e.g. by symmetry. 
For this reason it would be imperative to explore what the analysis would yield for
other possible Hamiltonians (i.e. other pairing mechanisms).\\
\\
• The writing style is somewhat colloquial and lengthy — in line with a comment, but not typical for
PRB articles. This could be fixed during a review process but would likely be an issue raised.\\
\\
• Despite the topical relation to an experiment, the insight that Tc depends on the apical oxygen
distance is not per se new to neither the theoretical nor the experimental literature on cuprates.
In summary, I see enough weaknesses of the submitted manuscript that I acknowledge the editorial
rejection as a justifiable decision. I therefore do not see myself in the position to overturning it. 
One can of course be of the opinion that any reasonable manuscript written to scientific standards (which the current one certainly is) should undergo review. 
However, this is not anymore the policy of PRB,
necessitated by the scarcity of available reviewers.\\
\\
Sincerely yours\\
\\
Titus Neupert\\
\\
Prof. Dr. Titus Neupert\\
Full Professor\\
Condensed Matter Theory\\
Phone +41 44 63 54800\\
titus.neupert@uzh.ch\\
\\
Zurich, October 25, 2025\\
\\
Page 1/1

\clearpage

\subsection*{Reply to Editorial Criticism by 
PRB Editorial Board Member Prof. T. Neupert}
\label{Reply_to_Neupert}

\noindent
Editor: 
\dots
What speaks in favor of the manuscript is that is presents a quantitative theoretical analysis on a
relevant experiment and an outstanding open problem in theoretical physics.\\
\textcolor{blue}{Authors (A): Our replies are inserted in blue color in the text of the letter.
Sofia, 15 November, 2025
}\\
\\
Critical points are:\\
\\
• Beyond the experiment that was analyzed, no testable predictions of the presented ‘theory’ have
been pointed out.\\
\textcolor{blue}{A: The experiment itself is the testable prediction of the theory in a sense of confirmation of the theory.
Moreover, there is a prediction: the small negative curvature not observed in the SJTM experiment shown in Fig.~2.
More testable predictions are presented implicitly in the text with proper citations but not enumerated:
\begin{enumerate}
\item First of all, one and the same set of parameters taken from the explanation of the Josephson voltage super-modulation should almost get to describe quantitatively the correlation between the shape of the Fermi contour and the critical temperature $T_c$.
\item If we take the slope of the $\overline{n}_p$ versus $\delta$ 
line depicted in \Fref{Fig:SJTM} we have to
estimate $t_{as}$ transfer amplitude between the Cu$4s$ orbital and the apical O$2p_z$ orbital.
This amplitude can be evaluated by tight-binding fit of the electron zone structure.
\item Considering $J_{sd}$ as a main interaction between electrons at least qualitatively \textit{s-d} Hamiltonian should describe the anisotropy of the ARPES line along the Fermi contour; so called hot-cold spots phenomenology. 
At least  order of magnitude evaluation should be acceptably predicted.
\item Last but not least, a cluster \textit{ab-initio} calculations must confirm that $J_{sd}$ is the most intensive
double-electron exchange amplitude exceeding 1~eV.
\end{enumerate}
*) Some initial studies in this direction have already been performed but the final implementation of this schedule should not be the case to published the present final result of the super-modulation of the Josephson voltage and Cooper pair density.
In the present manuscript, a BCS analysis of a contemporary significant experiment is performed, the word `theory’ does not deserve to be written in quotation marks;
it is a matter of professional etiquette.
Putting theory in quotation marks is diminshing end mocking, therefore it is offensive and inappropriate in academic communication. 
}\\
\\
• Ultimately, the connection between the theory and experiment is a near-linear dependence of the electron pair density on the apex distance. 
This is obtained while fitting the model parameters such that $T_c$ at the minimal and maximal apex distances found in experiments are matched.
In any theory, to lowest order, $T_c$ would depend linearly on any model parameter, unless a linear term is forbidden, e.g. by symmetry.
For this reason it would be imperative to explore what the analysis would yield for other possible Hamiltonians (i.e. other pairing mechanisms).\\
\textcolor{blue}{A: This should be some colloquial joke; unique straight line between 2 points.
It was written in the text and now emphasized in the amended version of the manuscript.
There are no fitting parameters in our \textit{s-d}-LCAO-BCS analysis of the slope of the linear dependence between super-fluid density and apical distance
$\md\overline{n}_p/\md \delta$, i.e. no linear dependence has been incorporated into the Hamiltonian (or the theory) initially.
Moreover, it is evident from Fig.~2 that this theory predicts small negative curvature not observed by the SJTM experiment and we have stated this explicitly.
And this precisely is the theoretical prediction mentioned above.\\
\\
The triumph of the theory is the fact that parameters taken from \textit{ab initio} band calculations predict with an excellent agreement the results of a SJTM experiment.\\
\\
As for the other Hamiltonians, they are not the subject of our study and it is not imperative for us to analyze them since our Hamiltonian explains the experimental results.
The same holds for colleagues working with the other Hamiltonians, it is not imperative for them to explore `our' (our in sense used in our study, it is introduced by Zener before the BCS epoch) or other Hamiltonians and they naturally do not do it.\\ \\
But anyway to reply to this `imperatively', with a mathematical precision we have to mention that it is imperative to explore what the analysis would yield for other possible Hamiltonians can not be implemented, this problem has no solution, i.e. other pairing mechanisms can not explain the numerical magnitude of the observed super-modulation of the Josephson effect.
Here we support the authors or Ref.~\cite{Omahony:22} entitling their article
\emph{On the electron pairing mechanism of cooper-oxide high temperature superconductor}
because they strongly believe that this experiment is crucial and will contribute to the final solution for the pairing amplitude.
We have only performed technical state-of-the-art-solution only using well known results from the electron-band and BCS theories.
We consider as quite naturally the theory supporting the crucial experiment to be presented in PRB.
It is necessary only to remove some misunderstanding created perhaps by uncleanly written manuscript.
Even in the abstract of the first version of the manuscript it has been written:\\
\\
`We conclude that the logarithmic derivative
$\mathcal Q_{n_p}\equiv (\delta/n_p)\,\md n_p/\md\delta$ is the most convenient
dimensionless parameter to solve the 
``concurrence of predictions from-strong-correlation theories
for hole-doped cuprates’’ ‘.\\
\\
The main purpose of our study is to calculate the dimensionless parameter which determines the slope of the experimental data line depicted in \Fref{Fig:SJTM}.
We repeat again, this master parameter is calculate from first principles, without any fitting procedure or fitting parameter.
We have reached an excellent accuracy better than 10\% and nothing more is necessary.
Now it is turn for other theorists to explain the crucial SJTM in an alternative way.
How the apical oxygen will be incorporated in RVB model, we have neither qualification nor imagination to make this; let it be done by the more doable and productive (with incomparably larger H-index).
Hence, imperatively following the editor imperative, we completely agree with:
``For this reason it would be imperative to explore what the analysis would yield for
other possible Hamiltonians (i.e. other pairing mechanisms).’’
This competition will be the real casting for the models and mechanisms explaining the high-$T_c$ superconductivity pairing mechanism.
But in order this imperative to be applied, those works have to be published, 
not rejected by indisputable refutation like our manuscript. 
An at least one reviewer per manuscript.
Or in other words, no such solution will be ever published with this kind of journal editors attitude.
In any case, the SJTM experiment~\cite{Omahony:22} 
will accelerate the development of condensed matter physics.
}\\
\\
• The writing style is somewhat colloquial and lengthy — in line with a comment, but not typical for PRB articles.
This could be fixed during a review process but would likely be an issue raised.\\
\textcolor{blue}{A: The problem of the mechanism of high-$T_c$ superconductivity is considered 
in many theoretical articles in PRB, but why is the problem still not solved?
The solution should be unique and untypical. 
A colloquium is a typical ingredient in the University education in the whole world.
By the way, one or both of us, can drop in Zurich to give a talk presenting our theory.
}\\
\\
• Despite the topical relation to an experiment, the insight that $T_c$ depends on the apical oxygen distance is not per se new to neither the theoretical nor the experimental literature on cuprates.\\
\textcolor{blue}{A: Definitely this idea is from the year of the discovery of high-$T_c$ superconductivity.
In order to emphasize that this is a long standing problem in front of high-$T_c$ superconductivity 
we cite the brilliant idea by J.~R\"{o}hler who concluded
that the strength of the \emph{spd} hybridization is mainly controlled by the length of the apex bond~\cite{Roehler:00,Roehler:00a}.
More than 20 years later, the SJTM experiment~\cite{Omahony:22} confirmed this idea.
In the present manuscript, we represent the indispensable theoretical explanation only.\\
\\
Lastly, we need some clarity on this statement, double negation means positive statement, i.e. that this insight is actually new, has the reviewer made a wrong statement?
}\\
\\
In summary, I see enough weaknesses of the submitted manuscript that I acknowledge the editorial rejection as a justifiable decision.
I therefore do not see myself in the position to overturning it. \\
\textcolor{blue}{A: Actually nothing necessary is overturned. 
Some misunderstandings are clearly analyzed in the our replies above. 
And the main is that we have not performed any fit,
we have performed state-of-the-art theoretical analysis.
Two editors can easily check that the solved problem is in the primary interest for condensed matter physics in general.
We have performed state-of-the-art calculations within standard theoretical constructions:
1) BCS theory, 2) Kondo-Zener exchange interaction and tight-binding interpolation of 
\textit{ab initio} calculated electron band theory; those ingredients have never disputed.
Any calculation or hypothesis could be in error that is why we have made available the programs performing of our numerical calculation.
If a reviewer cannot find any error -- this means that the crucial experiment~\cite{Omahony:22}
which reveals the electron pairing mechanism of copper-oxide high
temperature superconductivity has been successfully explained.
}\\
\\
One can of course be of the opinion that any reasonable manuscript written to scientific standards (which
the current one certainly is) should undergo review. 
However, this is not anymore the policy of PRB,
necessitated by the scarcity of available reviewers.\\
\textcolor{blue}{A: We perfectly understand the ecological catastrophe:
the calamity of manuscripts and the extinction of available reviewers or at least their entry in the IUCN Red List as endangered\footnote{\href{https://portals.iucn.org/library/sites/library/files/documents/RL-2001-001-2nd.pdf}{IUCN Red List
Categories And Criteria}}.
Nevertheless, if the authors of Ref~\cite{Omahony:22} really performed the crucial experiment, it is worthwhile to reconsider the misunderstanding analyzed in this reply.\\
\\
In conclusion: It does not matter what the present PRB business model is and the prejudice of its editors,
due to the ``outstanding open problem'' of the mechanism of high-$T_c$ superconductivity, at least one professional reviewer is an acceptable expenditure.
It is ridiculous a consideration of a manuscript to be terminated only because 
all critical points are in error (for instance ``fitting the model parameters'') or irrelevant (for instance the requirement for other Hamiltonians) to the study described in it.
}
\\
\\
\dots
in my role as PRB editorial board member \dots
I have carefully studied the manuscript, the relevant Ref.~1, and all
previous correspondence.\dots
This is obtained while fitting the model parameters such that $T_c$ at the minimal and maximal apex distances found in experiments are matched.
In any theory, to lowest order, \dots
any model parameter, \dots
For this reason it would be imperative to explore what the analysis would yield for other possible Hamiltonians (i.e. other pairing mechanisms).
\\
\textcolor{blue}{
Now we have to analyze this compound of incorrect statesmens:
If the manuscript was carefully studied, the editor would have observed that
there are no fitting parameter explaining the SJTM experiment.
All necessary parameters for our microscopic explanation
of the analyzed experiment are taken for \textit{ab initio}
electron band calculations and as a consequence the presented in the 
manuscript theory has the status of fundamental \textit{ab initio}
calculation without any fitting parameters taken from the explained experiment.
Led by perfect intuition, the authors of the SJTM experiment chose a theoretical 
notion in the title: ``On the electron pairing mechanism of copper-oxide high
temperature superconductivity''.
No doubts, even the PRB editorial board member confirms that 
in the manuscript is explained
``a relevant experiment and an outstanding open problem in theoretical physics''.
What type of an experiment can support a solution of a long standing theoretical 
problem?
This must be a crucial experiment having no alternative explanations whatsoever;
it is obvious for any physicist, for Tom, Dick and Harry, 
and every PRB editorial board member.
The imperative: 
``reason it would be imperative to explore what the analysis would yield for other possible Hamiltonians (i.e. other pairing mechanisms)''
has no reasons.
A crucial experiment has to have only one explanation.
How many alternatives has quantum mechanics explaining even
the spectrum of the hydrogen atom;
how many alternatives have Maxwell equations explaining propagation
of light?
How many alternative theoretical explanations are given
explaining the analyzed SJTM experiment?
To find an alternative explanation is an appropriate homework
of a  PRB editorial board member or a reviewer,
and we will be honoured to be reviewers of this homework.
In short, the mechanism of high-$T_c$ superconductivity in cuprates
is the most intensive double exchange amplitude in the
condensed matter physics.
The Zener \textit{s-d} exchange with Kondo antiferromagnetic sign.
The full Prof. Neupert suggestion
``for other possible Hamiltonians (i.e. other pairing mechanisms)''
is congenial to the
Mark Twain resume of an Italian ballad that all knights married the Princess: 
``The practiced knights from Palestine made holyday sport of carving 
the awkward men-at-arms into chops and steaks.  
The victory was complete.  Happiness reigned.  
The knights all married the daughter.  Joy! wassail! finis!''.
\\
}\\
• The writing style is somewhat colloquial and lengthy — in line with a comment, but not typical for
PRB articles. This could be fixed during a review process but would likely be an issue raised.\\
\textcolor{blue}{
Lastly, to comment on the not so harmless statement above.
This issue is very easily resolvable, the authors will give the manuscript text to any AI-text generator to transform the writing style to a ``colloquial'' one.
We even hope the editorial board member or any other PRB editor to suggest such an AI generator that gives an acceptable writing style for PRB.
}
\\
\\Yours sincerely,\\
\\
Todor Mishonov and Albert Varonov\\
16 Nov 2025


\clearpage

\section*{npj Quantum Materials}
\noindent
\\
\textbf{Critical remark on which editorial solution is based}:\\
\\
This work employs a LCAO model for curates that is solved in mean field theory. 
It does not connect to the modern material-specific correlated electron theory. 
As such, it may be an interesting phenomenological attempt, tuned by the use of additional Z-factors etc. to account for electron correlations and to explain experimental observations, 
Yet, by construction it remains an adhoc approach with many approximations and simplifications that will not gain enough credibility to obtain later impact. \\
\\
\textbf{Reply to the criticisms}:\\
\\
\textcolor{blue}
{First of all we wish to mention that 
the energy re-normalizing factors $Z_\epsilon$ and $Z_a$ are 
just to match the microscopic LDA approximations
with ARPES data; for our consideration these fitting parameters can be omitted.
}
The text in color is from copied from the amended version of the manuscript.\\
\\
The contemporary electron band theory does not need to be defended, this is the basis of our initial understanding of the electronic properties of high-$T_c$ cuprates. 
The electron band theory with LDA is the important initial point to start the studying of the highly correlated materials.
The electron wave function in muffin-tin approximation is constructed by a 
Linear Combination of Atomic-spheres Orbitals (LCAO) and the old abbreviation can be used as an alternative of the Tight Binding (TB) modeling of the electron band structure of CuO$_2$ plane superconductors.
The TB modeling is the realistic and an indispensable tool for creation of adequate lattice models for superconductivity.
Long time ago Phil Anderson characterized such TB models as 
``aesthetically attractive’’ and the basis for further construction of exchange Hamiltonians.
Here we reproduce a detail from the amended version of the manuscript
\textcolor{blue}
{
As it was emphasized by Anderson~\cite[p. 14, 15, 207]{Anderson_th}
LCAO or Tight Binding (TB) representation is intellectually viable and likely 
very accurate.
For the description of the conduction band we use the physics of
downswings used so successfully by O.~K.~Andersen~\cite[p. 15]{Anderson_th}.
}
Another American Nobel prize laureate Alexey Abrikosov pointed out one ``especially clear’’ explanation of TB modeling of high-$T_c$ superconductivity and used his article on the basis of his consideration of metal-insulator transition for undoped materials.
Here we copy another addendum from the edited version pf the manuscript
\textcolor{blue}
{In the present study we use the system of notations pointed out 
by Abrikosov~\cite{Abrikosov:03} as a \textit{very clear explanation} 
of the TB method applied to CuO$_2$ plane.
}
In order to reject a manuscript the referee entry in a contradiction with two
American Nobel prize laureates. 
A perfect demonstration how lack of physical intuition can be lucky
combined with complete ignoring of the literature on the subject.\\
\\
Metal-insulator transition is a phenomenon related to strong electron correlations.
Can it be created by
Kondo-Zener exchange interaction is another open problem for condensed matter physics.
We mentioned two Nobel prizes expressing an opinion opposite to the editor's remarks.
In order to alleviate further consideration of our manuscript, we suggest the following several questions with Yes/No reply.\\
\\
\textbf{Questions:}\\
\\
\begin{enumerate}
\item The experimental study by
Scanning Josephson Tunneling Microscopy (SJTM) 
revealed a new effect in the rich-effect physics of high-$T_c$ cuprates.
[On the electron pairing mechanism of copper-oxide high
temperature superconductivity,
by S.~M.~O’Mahony, Wangping~Ren, Weijiong~Chen, Yi~Xue~Chong, Xiaolong Liu, H.~Eisaki, S.~Uchida, M.~H.~Hamidian, and J.~C.~Séamus~Davis,
\href{https://www.pnas.org/doi/full/10.1073/pnas.2207449119}{PNAS, Vol.~\textbf{119}(37), e2207449119 (2022)}.
This is an experimental study using SJTM but the authors
are brave enough to point out in the title of their article
that they reveal the long searched pairing mechanism in the 
high-$T_c$ cuprates. 
\\
\\
Question: Can the solution of this important problem 
(electron pairing mechanism of copper-oxide high
temperature superconductivity)
be of interest for the readers of npj Quantum Materials? Yes/No.

\item Authors in the manuscript use a tight binding approximation of the contemporary 
atomic sphere band calculation together with \textit{s-d} double exchange interaction
which take into account strong electron correlation.
\\
\\
Question: Are there any alternative microscopic  explanation of the discovered
by SJTM super-modulation of the Josephson effect? Yes/No.\\
\\
\item The remark ``it may be an interesting phenomenological attempt'' is wrong and intentionally incorrect, which is in direct contradiction with the first sentence from the editor assessment ``This work employs a LCAO model for curates that is solved in mean field theory.''\\
\\
Question: Is this a ``phenomenological attempt''? Yes/No.\\
\\
\item Almost any theory includes ``approximations and simplifications'', moreover in such complex studies as strongly correlated systems.
And this cannot be an issue, as long as these ``approximations and simplifications'' are clearly argumented and appropriate for the specific problem.
If inappropriate, a peer review must point what and why it is inappropriate.\\
\\
Question: With these ``many approximations and simplifications'' (the authors do not agree with the word ``many'' since many means $\gg 1$), does the theoretical results from the manuscript quantitatively match with the obtained results from the SJTM experiment? Yes/No.
\end{enumerate}
\noindent
We consider that 4 bits additional referee review will be harmless to the business
model of the journal. 
That is why we are kindly waiting for a reply.\\
\\
Irrespectively to the prejudices and contemporaneity journal business model, 
we consider that a scientific content is the main ingredient for a manuscript
submitted to an academic journal. 
One manuscript cannot be rejected by completely wrong arguments.\\
\\
Truly yours,\\
\\
Todor Mishonov and Albert Varonov

\end{document}